\documentclass[journal,draftclsnofoot, 12pt,onecolumn]{IEEEtran}
\ifCLASSINFOpdf
\else
   \usepackage[dvips]{graphicx}
\fi
\usepackage{url}
\usepackage{graphicx}
\usepackage{amsmath,bm,bbm}
\usepackage{amssymb}  
\usepackage{cite}        
 \usepackage{epstopdf} 
  \usepackage{xcolor} 
  \usepackage{subfigure} 
\usepackage{amsthm}    
\usepackage{booktabs}
\usepackage{multirow}
\usepackage{makecell}  
\usepackage{enumerate}
\usepackage{algorithm}
\usepackage{algpseudocode}
\usepackage{stfloats}
\usepackage{cuted}

\usepackage{setspace}
\allowdisplaybreaks[4]
\begin{document}

\title{Federated Dynamic Neural Network for Deep MIMO Detection}

\author{Yuwen Yang,  Feifei Gao, Jiang Xue, and Ting Zhou,  and Zongben Xu    
\thanks{Y. Yang and F. Gao are with  Institute for Artificial Intelligence Tsinghua University
(THUAI), State Key Lab of Intelligent Technologies and Systems, Beijing National Research Center for Information Science and
Technology (BNRist), Department of Automation, Tsinghua University, Beijing,
100084, P. R. China (email: yyw18@mails.tsinghua.edu.cn, feifeigao@ieee.org).}
\thanks{J. Xue and Z. Xu are with the National Engineering Laboratory for Big Data Analytics, Xi'an International Academy for Mathematics and Mathematical Technology,  School of Mathematics and Statistics, Xi'an Jiaotong University, Xi'an 710049, P. R. China
(email:  x.jiang@xjtu.edu.cn; zbxu@xjtu.edu.cn).}
\thanks{T. Zhou is with the Shanghai Advanced Research Institute, CAS, Shanghai Frontier Innovation and Research Institute, Shanghai 201210, P.R. China (e-mail: zhouting@sari.ac.cn). }
}

\markboth{XXXX, VOL. XX, NO. XX, XXX }
{Shell \MakeLowercase{\textit{et al.}}: xxxx}
\maketitle

\begin{abstract}
In this paper, we develop a dynamic detection network (DDNet)
based detector for multiple-input multiple-output  (MIMO) systems.
By constructing an improved DetNet (IDetNet) detector and the OAMPNet detector  as two
independent network branches, the DDNet detector  performs sample-wise dynamic routing to adaptively select a better one between the IDetNet and the OAMPNet detectors  for every  samples under different system conditions.
To avoid the  prohibitive transmission  overhead of dataset collection in  centralized learning (CL),
we propose the federated averaging (FedAve)-DDNet detector,  where  all  raw data  are kept at local clients and only locally trained model parameters are transmitted to the central server for aggregation.  To further reduce the transmission overhead, we develop the federated gradient sparsification (FedGS)-DDNet detector by randomly sampling gradients with elaborately calculated probability when uploading gradients to the central server.
Based on simulation results,   the proposed DDNet detector consistently outperforms other detectors  under all system conditions thanks to the  sample-wise dynamic routing. Moreover,   the federated DDNet  detectors, especially the FedGS-DDNet detector, can reduce the transmission overhead  by at least 25.7\%  while maintaining  satisfactory detection accuracy.
\end{abstract}
\vspace{-8mm}


\IEEEpeerreviewmaketitle

\section{Introduction}

Multiple-input multiple-output (MIMO) technique  has been widely applied to various wireless communication systems for its high spectrum  efficiency and link reliability \cite{8443598,8354789}.  To embrace these benefits, efficient
signal detection algorithms are  critical to the receiver design \cite{4815548}.
Over the last decades, various  detectors with different complexities have been proposed, among which the maximum likehood (ML) detector can obtain the best accuracy by exhaustively searching the possible signal space.
However, the overwhelming complexity of the ML detector makes it nearly impractical  in real systems.
Sphere decoding (SD) algorithm \cite{1194444}  limits the searching space  and can achieve near-optimal  performance.
Other detectors that offer relatively desired  performance with lower  complexity include  the approximate message passing (AMP) detector \cite{6778065} and the semidefinite relaxation (SDR) detector \cite{5447068}. The AMP  detector is derived from a strict approximation of Gaussian belief propagation and performs well on independent  identically distributed (i.i.d.) Gaussian channels. The SDR detector formulates  the MIMO detection problem as a non-convex homogeneous quadratically constrained quadratic problem and exhibits strong robustness.
Besides, the linear minimum-mean-squared-error (LMMSE) detector provides  acceptable  performance with lower accuracy and is   widely adopted  in practical systems \cite{7244171}.

Motivated by the successful  applications  of deep learning (DL) in physical layer communications \cite{8353153,8752012,8322184,8795533,8672767,alrabeiah2019deep,8922743,9288911,hu2021understanding},
many recent works resort to DL, especially deep unfolding for the  MIMO detection problem \cite{8642915,9018199,8052521,9159940,9075976,8936847,8780959}.
In deep unfolding, trainable weights and non-linearities are added to each iteration of the detection, and  are then  optimized to improve the  performance.
For example,the DetNet detector  \cite{8642915}  unfolds  a projected gradient descent algorithm, which can achieve better accuracy  than  the  SDR detector in most cases.
 By unfolding   the orthogonal AMP (OAMP) algorithm and adding only several trainable parameters per layer, the OAMPNet detector \cite{9018199} exhibits  promising advantages over the conventional OAMP algorithm.
Despite various   deep unfolding-based detectors were subsequently proposed  \cite{9159940,9075976,8936847,8780959},  there is no  single detector that can achieve
the optimal accuracy  with acceptable complexity under all different system setups and channel conditions.

Recently, dynamic neural networks have attracted growing attention for its remarkable
advantages in terms of adaptiveness, accuracy, and computational efficiency   \cite{han2021dynamic,Hinton10,fedus2021switch}. Most prevalent DL algorithms \cite{9277535,guo2019convolutional,9175003,9136588,weihua20192d,9076084,8663966,8985539} adopt static  models, where
 both the  computational structures and the network parameters are fixed in the testing stage, i.e., every testing input goes through the same mathematical  operations, which may limit the representation power and the efficiency of the DL models \cite{Hinton10}.
In contrast,  dynamic neural networks adapt their structures or parameters to different inputs by selectively activating model components, e.g., layers  or subnetworks, according to different inputs. For instance, for the switch transformer structure in  \cite{fedus2021switch},  multiple network branches are built in parallel and are selectively  executed based on the prediction of a front routing layers. In particular, only the $K$ network branches corresponding to the top-$K$ elements of the routing vector would be activated by the  front routing layers, and an auxiliary loss is adopted to encourage these network branches to be activated with a balanced probability.
The  dynamic structure proposed in \cite{fedus2021switch} illuminates a potential solution to   MIMO detection    under varying system conditions. More specifically, we can build different detectors as independent network branches  and design a routing module to adaptively select an optimal detector conditioned on the  input samples that contain the information of the system conditions.

On the other hand, most DL based works for  physical layer communications are based on {centralized learning} (CL), where the networks are trained in the central server with the training data collected from  the clients. {However, the  transmission of the whole training dataset  from the clients to the central server  results in prohibitive transmission overhead and  poses a threat to data privacy.
To tackle  these problems, \emph{decentralized learning} becomes a natural  solution,  where all the raw data are kept at clients and only  locally trained model parameters are transmitted to the central server   \cite{9141214,mcmahan2017nt,li2018federated,9500877}.}
Another underlying motivation is that the local training enables  decentralized learning algorithms to handle the real-time changes lying in local datasets, which is a  competitive advantage for  latency sensitive applications like MIMO detection in dynamic  wireless communication environments  \cite{9141214}.
 As an emerging branch of decentralized learning, federated learning (FL)    distinguishes    from  other  decentralized learning approaches in that it can deal with non-i.i.d. and unbalanced local datasets \cite{li2018federated,mcmahan2017nt,9500877}.  There are already some works that leverage FL  for wireless communication applications \cite{qin2020federated}, such  as channel estimation \cite{elbir2020federated}, hybrid beamforming \cite{9177084}, and intelligent reflecting surface (IRS) achievable rate optimization \cite{9145388}, etc.
In particular,   FL based channel estimation algorithms have been  developed in \cite{elbir2020federated}
for both conventional and IRS assisted massive MIMO  systems, which   obtain satisfactory performance closed to  CL but require much lower transmission overhead.
In \cite{9177084}, a FL based hybrid beamforming  scheme has been  proposed  for mmwave massive MIMO systems.
Notice that in both \cite{elbir2020federated} and \cite{9177084}, the clients only conduct one time stochastic gradient descent and then  upload  the gradients to the  central server during every epoch, which is sometimes  inefficient
\cite{zhou2020communication}.  To enhance the training efficiency, a federated averaging (FedAve) algorithm in \cite{mcmahan2017nt} has been developed, where  clients   execute multiple local parameter updates before  uploading  the network weights to the  central server. The work in \cite{li2019convergence} has provided the theoretical analysis of FedAve algorithm and proved its  convergence  on both i.i.d. and non-i.i.d. data with decaying learning
rate.  To further reduce the transmission overhead, \cite{lin2018deep} and \cite{Wangni10}   drop out small gradients when transmitting locally trained gradients to the central server.

To improve the detection accuracy, reduce the transmission overhead, and protect the  data privacy,   we develop the federated \textbf{d}ynamic  \textbf{d}etection \textbf{net}work (DDNet) based detectors for  MIMO systems by  borrowing ideas from  dynamic neural networks and  FL.
The main contributions of this work can be summarized as following:
\begin{itemize}
 \item  {To enhance the overall  detection accuracy  under varying system conditions, we design the  architecture of the DDNet detector, where an \textbf{i}mproved \textbf{DetNet} (IDetNet) detector and the OAMPNet detector are built as two independent network branches. Moreover, a specially designed \textbf{route} \textbf{net}work (RouteNet) performs sample-wise dynamic routing among the IDetNet and the OAMPNet detectors, i.e., adaptively  selecting  a better detector for every sample under different system conditions. To the best of authors' knowledge, this is the first work that introduces  dynamic  neural networks into wireless communications.}
 {\item To reduce the transmission overhead and protect the  data privacy,  we propose the  FedAve-DDNet detector, where   OAMPNet   is trained in CL way by the clients for its low training cost, while  IDetNet and RouteNet are successively trained by the FedAve algorithm.
  \item To further reduce the transmission overhead, we develop the federated gradient sparsification (FedGS)-DDNet detector by randomly discarding  gradients with elaborately calculated probability  while uploading local gradients to the central server. The gradient sparsification  technique only involves    addition, multiplication, and minimization operations, and therefore is computationally efficient.}
\end{itemize}

The rest of this paper is organized as follows. The
  MIMO detection problem is formulated in Section \ref{secmodelsdf}.
  The architecture of the DDNet detector  is  presented in Section \ref{secsddd}. The FedAve-DDNet and the FedGS-DDNet detectors are  developed in Section \ref{secfddse}.
Numerical results are provided in Section \ref{secnumer}, and  main conclusions are
given in Section \ref{secconcul}.

\emph{Notations:}
The bold and lowercase letters denote vectors while the bold and
capital letters denote matrices; $[ \bm z ]_{p}$ and  $\textrm{len}(\bm z)$ denote the $p$-th entry and the length of the vector $\bm z$, respectively;  $\Re[\cdot]$ and $\Im[\cdot]$, respectively,  denote  the real and the imaginary parts of  matrices, vectors, or  scales; $\left\| {\bm{x}} \right\|_1$  and $\left\| {\bm{x}} \right\|_2$ respectively denote the $L_1$  and the $L_2$  norms of  $\bm x$; $|\mathbb{D}|$ denote the number of elements in the dataset $\mathbb{D}$; $(\cdot)^T$  denotes the transpose of a matrix or a vector;  $\textrm{tr}(\cdot)$ and $\textrm{vec}(\cdot)$    denote  the trace and the  vectorization  of a matrix, respectively; ${\mathbb{C}^{m \times n}}$ represents the $m \times n$  complex vector  space; $ \circ $  represents the composite mapping operation; $ \mathcal{N}_{C}(\bm{0,I})$ and $ \mathcal{N}(\bm{0,I})$ respectively represent the standard complex and real Gaussian distributions; $ E [\cdot]$  represents the expectation with respect to all random variables within the brackets;  $\leftarrow $ represents the  assignment operation.

\section{Problem Formulation}\label{secmodelsdf}
We consider a general MIMO system as the following: 
\begin{equation}\label{equmimo}
\tilde{\bm{y}}=\tilde{\bm{H}} \tilde{\bm{x}}+\tilde{\bm{n}},
\end{equation}
where  $\tilde{\bm{y}} \in \mathbb{C}^{N_r \times 1} $ is the received signal vector,
$\tilde{\bm{H}} \in \mathbb{C}^{N_r \times N_t} $ is the channel matrix, $\tilde{\bm{x}}$ is the transmitted symbol vector drawn from the constellation alphabet $\mathbb{A}$, and $\tilde{\bm{n}} \in  \mathcal{N}_{C}({0,\sigma_{n}^2 \bm I})$
is the  Gaussian noise with variance $\sigma_{n}^2$.
We  transform Eq.~\eqref{equmimo} into the real domain as ${\bm{y}}={\bm{H}} {\bm{x}}+{\bm{n}},$
where  ${\bm{y}}=[\Re[\bm{y}],\Im[\bm{y}] ]^T$, ${\bm{x}}=[\Re[\bm{x}],\Im[\bm{x}] ]^T$,
 ${\bm{n}}=[\Re[\bm{n}],\Im[\bm{n}] ]^T$, and
 \begin{equation}\label{equre}
 \bm{H} \triangleq\left[\begin{array}{cc}
\Re(\tilde{\bm{H}}) & -\Im(\tilde{\bm{H}}) \\
\Im(\tilde{\bm{H}}) & \Re(\tilde{\bm{H}})
\end{array}\right].
\end{equation}
The aim of  detection algorithms is to recover the transmitted signal vector ${\bm{x}}$ from the received signal vector ${\bm{y}}$ given a known channel matrix $\bm{H}$ at the receiver.
The LMMSE detector \cite{7244171} is one of the most widely adopted ones  and can be expressed as
 \begin{equation}\label{equlmsse}
\hat{\bm{x}}_{\textrm{LMMSE}}=\mathcal{Q}\left[\left(\bm{H}^{T} \bm{H}+ \sigma_{n}^2 \bm I\right)^{-1} \bm{H}^{T} \bm{y}\right],
\end{equation}
where $\mathcal{Q}[\cdot]$ is the quantizer associated with the constellation alphabet  $\mathbb{A}$.



\section{Dynamic Detection Network}\label{secsddd}
\begin{figure}[!t]
\centering
\includegraphics[width=90mm]{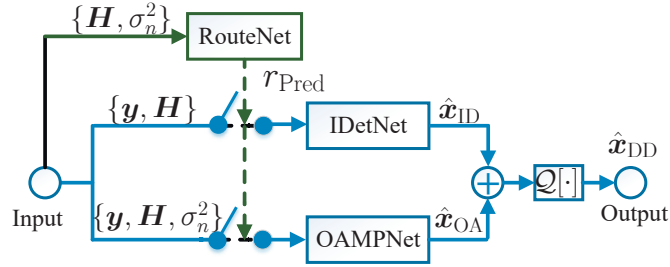}
\caption{The structure of DDNet.}
\label{figmoe}
\end{figure}
As illustrated in Fig.~\ref{figmoe}, the proposed DDNet consists of three subnetworks, i.e., IDetNet, OAMPNet and the routing module RouteNet. Note that  IDetNet and OAMPNet, i.e.,  two different detectors, are built as two parallel network branches that are conditionally executed based on the predictions of RouteNet.
The output of DDNet  can be written as 
{ \begin{equation}\label{equend}
\hat{\bm x}_{\textrm{DD}}=\mathcal{Q}[(1-r_{\textrm{Pred}})\hat{\bm x}_{\textrm{ID}}+r_{\textrm{Pred}}\hat{\bm x}_{\textrm{OA}}],
\end{equation}
where $r_{\textrm{Pred}}\in\{0,1\}$  is the  route index predicted by RouteNet, $\hat{\bm x}_{\textrm{ID}}$ is the estimate  of IDetNet, and $\hat{\bm x}_{\textrm{OA}}$ is the estimate  of OAMPNet. As indicated by Eq.~\eqref{equend},
only one of  IDetNet and OAMPNet would be activated by RouteNet for each sample. When the detection accuracies of IDetNet and OAMPNet for one sample are different, RouteNet would select
the detector with higher accuracy. Otherwise, RouteNet would select the detector with lower complexity.}

In the following, we will provide detailed descriptions of the three  subnetworks,   followed by detailed CL training steps and the complexity analysis.

\subsection{IDetNet}\label{secidetnet} 
The structure of IDetNet  is enlightened  by \cite{8642915}.
Each layer of  IDetNet mimics one iteration of the  projected gradient descent optimization, and the  $k$-th layer  can be mathematically expressed as
\begin{subequations}
\begin{eqnarray}
  \bm z_{k} \!\!&=&\!\! {\mathcal{ F}}_{\mathrm{Relu}}({\mathcal{ F}}_{\mathrm{Den}}^{k,1}([\bm v_k, \bm H^T\bm y, \bm H^T\bm H \hat{\bm x}_k, \hat{\bm x}_k ])), \\
  {\bm v}_{k+1} \!\!&=&\!\! \mathcal{F}_{\mathrm{Smo}}^{k,1}({\mathcal{ F}}_{\mathrm{Den}}^{k,2}(\bm z_{k}),{\bm v}_{k}), \\
\hat{\bm x}_{k+1} \!\!&=&\!\! \mathcal{F}_{\mathrm{Smo}}^{k,2}({\mathcal{ F}}_{\mathrm{Lss}}^{k}({\mathcal{ F}}_{\mathrm{Den}}^{k,3}(\bm z_{k})),\hat{\bm x}_{k}),
\end{eqnarray}
\end{subequations}
where both $\bm z_{k}$ and $\bm v_{k}$ are intermediate iteration variables,  ${\mathcal{ F}}_{\mathrm{Relu}}(\bm s) = \max\{\bm s, 0\}$ is the nonlinear activation function, and
 \begin{equation}\label{equlss}
{\mathcal{ F}}_{\mathrm{Lss}}^{k}(\bm s; \beta_k) = -1+ \frac{{\mathcal{ F}}_{\mathrm{Relu}}(\bm s+\beta_k) }{|\beta_k|}-\frac{{\mathcal{ F}}_{\mathrm{Relu}}(\bm s-\beta_k) }{|\beta_k|}
\end{equation}
is the element-wise linear soft sign function with $\beta_k$ being a trainable parameter.
Besides, the $i$-th dense layer in the $k$-th layer of IDetNet can be expressed as
 \begin{equation}\label{equden}
\mathcal{F}_{\mathrm{Den}}^{k,i} (\bm s; \bm w_{k,i},\bm b_{k,i})= \bm w_{k,i}\bm s+\bm b_{k,i},  \quad i=1,2,3,
\end{equation}
where $\bm w_{k,i}$ and $\bm b_{k,i}$ are respectively the weight and the bias of the dense layer.
Moreover, we adopt the following smoothing function
 \begin{equation}\label{equsmo}
\mathcal{F}_{\mathrm{Smo}}^{k,i}(\bm s_{{k+1}},\bm s_{{k}}, \alpha_{k,i})= (1-\alpha_{k,i})\bm s_{{k+1}} +\alpha_{k,i}\bm s_{{k}}, \ i=1,2,
\end{equation}
where $\alpha_{k,i}$ is the trainable smoothing factor.
Denote $K_{\textrm{ID}}$  as the layer number of IDetNet.
By cascading  $K_{\textrm{ID}}$  layers, we   can obtain the output of IDetNet as
  \begin{equation}\label{equxkid}
\hat{\bm x}_{\textrm{ID}}\buildrel \Delta \over =\hat{\bm x}_{K_{\textrm{ID}}+1} \buildrel \Delta \over =\mathcal{F}_{\mathrm{IDetNet}}(\bm H, \bm y; \bm \Omega_{\mathrm{ID}}),
\end{equation}
 where $\bm \Omega_{\mathrm{ID}}\buildrel \Delta \over =\{ \beta_k,\{\bm w_{k,i},\bm b_{k,i}\}_{i=1,2,3},\{\alpha_{k,i}\}_{i=1,2}\}_{k=1}^{K_{\textrm{ID}}}$ is the trainable parameter  of IDetNet to be optimized.
 The loss function of IDetNet can be written as
 \begin{equation}\label{equloss}
   \mathcal{L}_{\textrm{ID}}\left(\bm \Omega_{\mathrm{ID}}\right) = \frac{1}{{D}}\sum\limits_{d=1}^{D}
\sum\limits_{k= 2}^{K_{\textrm{ID}}+1} \|\hat{\bm x}_{k}^{(d)}-\bm x^{(d)}\|_{2}^{2},
\end{equation}
where $D$ is the batch size, and
  $d$ denotes the index of the   training samples.

Compared with DetNet in \cite{8642915}, we make two improvements in IDetNet:
\begin{enumerate}[1)]
  \item We set  $\{\beta_k\}_{k=1}^{K_{\textrm{ID}}}$  to be  trainable parameters, which allows  each layer of IDetNet to have a linear soft sign function with different softness. As indicated in Eq.~\eqref{equlss},  the function ${\mathcal{ F}}_{\mathrm{Lss}}^{k}$ with a smaller $\beta_k$ has a larger slope near the zero point\footnote{The curves of ${\mathcal{ F}}_{\mathrm{Lss}}^{k}$ versus varying $\beta_k$ can be found in \cite{8642915}.}  and thus makes a harder decision about the input $\bm s$, which  results in faster convergence but may potentially increase the accumulated estimated error of the subsequent layers.
        In IDetNet, layers at different depths are able to learn proper $\{\beta_k\}_{k=1}^{K_{\textrm{ID}}}$  to improve the convergence speed and achieve better performance. 
  \item We exploit   smoothing function $\mathcal{F}_{\mathrm{Smo}}^{k,i}$ for the updates of $\hat{\bm x}_{k}$ and $\bm v_{k}$. The smoothing factor, $\alpha_{k,i}$, is a trainable parameters, and allows each layer of IDetNet to assign different weights to the outputs of the prior layer. The smoothing function can improve the stability and accuracy of the network.
\end{enumerate}


\subsection{OAMPNet}
OAMPNet,   originally proposed in  \cite{9018199}, will be briefly illustrated here.
The  $k$-th layer of OAMPNet   can be described as following:
\begin{subequations}
\begin{align}
v_{k}^2 &= \frac{\|\bm y -\bm H \hat{\bm x}_k\|_{2}^{2}-\textrm{tr}(\bm R_{\bm n})}{\textrm{tr}(\bm H^T\bm H)}, \\
\bm A_{k} &=  \frac{2N_t v_{k}^2 \bm H^T( v_{k}^2\bm{HH}^T+\bm R_{\bm n})^{-1}}{\textrm{tr}(v_{k}^2 \bm H^T( v_{k}^2\bm{HH}^T+\bm R_{\bm n})^{-1}\bm H)}, \label{equbsd}\\
  \bm z_{k}  &=  \hat{\bm x}_k+\gamma_{k,1}\bm A_{k}(\bm y -\bm H \hat{\bm x}_k), \\
\bm C_{k} &= \bm I -\gamma_{k,2}\bm A_{k}\bm H , \\
\tau_{k}^{2} &= \frac{{\textrm{tr}(\bm C_{k} \bm C_{k}^T)}v_{k}^2+{\textrm{tr}(\bm A_{k}\bm R_{\bm n} \bm A_{k}^T)}}{2N_t},  \\
\hat{\bm x}_{ k+1} &=  \eta_{k}\left(\bm z_{k}, \tau_{k}^{2} ; \gamma_{k,3}, \gamma_{k,4}\right),
\end{align}
\end{subequations}
where $\bm R_{\bm n}=\sigma_{n}^2\bm I$ is the covariance matrix of the noise $\bm n$, $v_{k}^2$ and $\tau_{k}^{2}$ are error estimators, $\bm A_{k}$ is the de-correlated matrix, $\bm z_{k}$ is the linear estimator,  and $\bm C_{k}$ is an intermediate iteration variable. Moreover, $\eta_{k}(\cdot)$ is the nonlinear estimator, expressed as 
\begin{equation}\label{equxkid}
\eta_{k}\left(\bm z_{k}, \tau_{k}^{2} ; \gamma_{k,3}, \gamma_{k,4}\right)\! =\!\gamma_{k,3}\left({E} \left\{\bm{x} | \bm z_{k}, \tau_{k}^2\right\}-\gamma_{k,4} \bm z_{k}\right),
\end{equation}
where ${E} \left\{\bm{x} | \bm z_{k}, \tau_{k}^2\right\}$ is the MMSE estimator of $\bm x$.
Denote $K_{\textrm{OA}}$  as the  number of layers in OAMPNet.
 The output of OAMPNet can be expressed as
  \begin{equation}\label{equxkoa}
\hat{\bm x}_{\textrm{OA}}\buildrel \Delta \over =\hat{\bm x}_{K_{\textrm{OA}}+1} \buildrel \Delta \over =\mathcal{F}_{\mathrm{OAMPNet}}(\bm H, \bm y, \sigma_{n}^2; \bm \Omega_{\mathrm{OA}}),
\end{equation}
where
$\bm \Omega_{\mathrm{OA}}\buildrel \Delta \over = \{\gamma_{i,k}\}_{i,k=1}^{4,K_{\textrm{OA}}}$
is the trainable parameter set of OAMPNet to be optimized.
 The loss function of OAMPNet is
 \begin{equation}\label{equlosoap}
   \mathcal{L}_{\textrm{OA}}\left(\bm \Omega_{\mathrm{OA}}\right) = \frac{1}{{D}}\sum\limits_{d=1}^{D}
\sum\limits_{k= 2}^{K_{\textrm{OA}}+1} \|\hat{\bm x}_{k}^{(d)}-\bm x^{(d)}\|_{2}^{2}.
\end{equation}

\begin{figure}[!t]
\centering
\includegraphics[width=90mm]{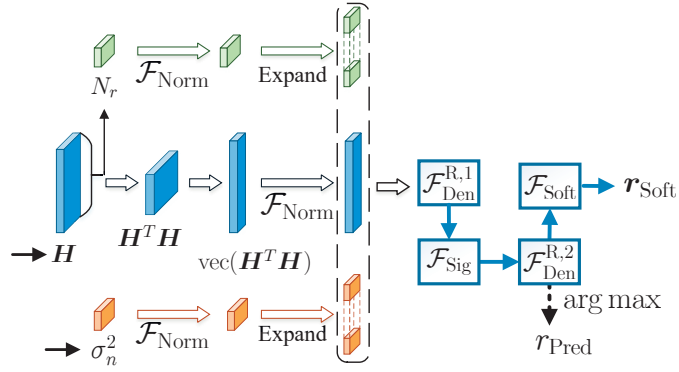}
\caption{The structure of RouteNet.}
\label{figroute}
\end{figure}

\subsection{RouteNet}\label{secroutend}
The goal of RouteNet is to perform sample-wise dynamic routing between   IDetNet and OAMPNet, i.e., adaptively selecting a better subnetwork for each sample according to the system conditions, including the signal-to-noise ratio (SNR), the number of received antennas, and the channel statistics. Therefore, we set the noise power $\sigma_{n}^{2}$, the number of received antennas $N_r$, and the channel matrix  $\bm H$ as the input data to instruct  RouteNet in subnetwork routing under different system conditions.
{As shown in Fig.~\ref{figroute}, we first infer the value of  $N_r$ from the shape of   $\bm H$ for each sample, and then adopt $\bm H^T\bm H$ rather than  $\bm H$  as an input
to   avoid the varying dimensionality of $\bm H$ in input samples.}
Moreover, we normalize $\sigma_{n}^{2}$,  $N_r$, and $\bm H$ to a unified interval by a linear transformation,
\begin{equation}\label{equxnor}
{\mathcal{ F}}_{\mathrm{Norm}}(\bm s; \bm s_{\textrm{max}},\bm s_{\textrm{min}}) = (\bm s- \bm s_{\textrm{min}})/( \bm s_{\textrm{max}}-\bm s_{\textrm{min}}),
\end{equation}
where $\bm s_{\textrm{max}} \!=\! \max_{\bm s\in \mathbb{D}} \bm s$ and $\bm s_{\textrm{min}}\!= \! \min_{\bm s\in \mathbb{D}} \bm s$ are respectively the element-wise maximum  and minimum vectors  over datasets $\mathbb{D}$.
Next, we expand both ${\mathcal{ F}}_{\mathrm{Norm}}(\sigma_{n}^{2})$ and  ${\mathcal{ F}}_{\mathrm{Norm}} (N_r)$ to vectors of length  $N_t$. The reason that we conduct the dimension expansion   is to prevent ${\mathcal{ F}}_{\mathrm{Norm}}(\sigma_{n}^{2})$ and  ${\mathcal{ F}}_{\mathrm{Norm}} (N_r)$  from being ignored by RouteNet due to their extremely small dimensions. It is also reasonable to expand  $\sigma_{n}^{2}$ and  $N_r$ to vectors of other proper lengths.

As shown in Fig.~\ref{figroute}, the  input vector  of  RouteNet can be written as
\begin{equation}\label{equinfr}
\bm s_{\textrm{RO}}=[{\mathcal{ F}}_{\mathrm{Norm}}(\sigma_{n}^{2})\bm 1_{N_t},{\mathcal{ F}}_{\mathrm{Norm}}\textrm{vec}(\bm H^T\bm H) , {\mathcal{ F}}_{\mathrm{Norm}} (N_r)\bm 1_{N_t}].
\end{equation}
Since RouteNet only consists of two dense layers, i.e., $\{\mathcal{F}_{\mathrm{Den}}^{\textrm{R},i}\}_{i=1,2}$, we adopt the Sigmoid function, $\mathcal{F}_{\mathrm{Sig}}(\bm s)= 1/ (1+e^{-\bm s})$, instead of the ReLU function to improve the nonlinear fitting capability of  RouteNet.
We add the softmax function,  $\mathcal{F}_{\mathrm{Soft}}$, to the output layer, which is given by
\begin{equation}\label{equsd}
 \left[\mathcal{F}_{\mathrm{Soft}}(\bm s)\right]_{p} =\frac{e^{\left[\bm s\right]_{p} }}{\sum_{p=1}^{p=\textrm{len}(\bm s)}e^{\left[\bm s\right]_{p} }}, \ p= 1,2,\cdots,\textrm{len}(\bm s).
\end{equation}
{The route index $r_{\textrm{Pred}}\in\{0,1\}$
 can be obtained by
  \begin{eqnarray}
r_{\textrm{Pred}} =  \arg\max \{ \mathcal{F}_{\mathrm{Den}}^{\textrm{R},2} \circ {\mathcal{ F}}_{\mathrm{Sig}} \circ \mathcal{F}_{\mathrm{Den}}^{\textrm{R},1}(\bm s_{\textrm{RO}})\}-1.
\end{eqnarray}}
Define the  bit error function as
\begin{equation}\label{equacc}
{\mathcal{ F}}_{\mathrm{BE}}(\hat{\bm x};\bm x)= {\|\mathcal{Q}[\bm x]-\mathcal{Q}[\hat{\bm x}]\|_1},
\end{equation}
where  $\bm x$ and $\hat{\bm x}$ respectively  denote the transmitted signal and the estimated signal.
{The  route label $\bm r_{\textrm{Lab}}$ is obtained by measuring the  bit errors of OAMPNet and IDetNet, and can be written as
\begin{align}\label{equdlabgg}
\begin{array}{l}
\bm r_{\textrm{Lab}} = \left\{
\begin{array}{ll}
$[1,0]$,\quad & \textrm{if }{\mathcal{ F}}_{\mathrm{BE}}(\hat{\bm x}_{\textrm{ID}}) \leq {\mathcal{ F}}_{\mathrm{BE}}(\hat{\bm x}_{\textrm{OA}});\\
$[0,1]$, \quad & \mathrm{otherwise}.
\end{array} \right.
\end{array}
\end{align}
In the case of ${\mathcal{ F}}_{\mathrm{BE}}(\hat{\bm x}_{\textrm{ID}}) = {\mathcal{ F}}_{\mathrm{BE}}(\hat{\bm x}_{\textrm{OA}})$, IDetNet would be selected for its lower complexity compared with OAMPNet. The classification error in the case of ${\mathcal{ F}}_{\mathrm{BE}}(\hat{\bm x}_{\textrm{ID}}) = {\mathcal{ F}}_{\mathrm{BE}}(\hat{\bm x}_{\textrm{OA}})$ would increase the computational cost while the classification error in the case of ${\mathcal{ F}}_{\mathrm{BE}}(\hat{\bm x}_{\textrm{ID}}) \neq {\mathcal{ F}}_{\mathrm{BE}}(\hat{\bm x}_{\textrm{OA}})$ would increase the bit errors.
The loss function  can be written as
\begin{equation}\label{sfffgsfgs}
   \mathcal{L}_{\textrm{RO}}\left(\bm \Omega_{\textrm{RO}}\right) = \frac{1}{{D}}\sum\limits_{d=1}^{D}\left( entropy(\bm r_{\textrm{Lab}}^{(d)},\bm r_{\textrm{Soft}}^{(d)})
+\xi ( {\mathcal{ F}}_{\mathrm{BE}}(\hat{\bm x}_{\textrm{DD}}^{(d)}) - \min \{ {\mathcal{ F}}_{\mathrm{BE}}(\hat{\bm x}_{\textrm{ID}}^{(d)}), {\mathcal{ F}}_{\mathrm{BE}}(\hat{\bm x}_{\textrm{OA}}^{(d)})\})\right),
\end{equation}
where $entropy(\bm r_{\textrm{Lab}}^{(d)},\bm r_{\textrm{Soft}}^{(d)})= \sum\nolimits_{p= 1}^{ \textrm{len}(\bm r_{\textrm{Soft}})} \left[\bm r_{\textrm{Lab}}^{(d)}\right]_{p}\log \left[\bm r_{\textrm{Soft}}^{(d)}\right]_{p}$ is the categorical cross entropy, $\bm \Omega_{\textrm{RO}}\buildrel \Delta \over =\{\bm w_{\textrm{R},i},\bm b_{\textrm{R},i}\}_{i=1,2}$ is the trainable parameter of RouteNet,
$\bm r_{\textrm{Soft}} = {\mathcal{ F}}_{\mathrm{Soft}}\circ \mathcal{F}_{\mathrm{Den}}^{\textrm{R},2} \circ {\mathcal{ F}}_{\mathrm{Sig}} \circ \mathcal{F}_{\mathrm{Den}}^{\textrm{R},1}(\bm s_{\textrm{RO}})$  is the output of RouteNet, and
$\xi>0$ is the penalty coefficient. A larger $\xi$ indicates less tolerance to the accuracy loss of DDNet, and thus forces the algorithm to be more  sensitive to the classification errors that  increase the  bit errors.}

\emph{Remark 1:} Due to the high detection accuracy of the  IDetNet and the OAMPNet detectors in high SNR regions, most samples would be detected by the  IDetNet or the OAMPNet detectors  without errors. In this case, the number of  samples with the route label $\bm r_{\textrm{Lab}}=[1,0]$ would be significantly more than those with the route label $\bm r_{\textrm{Lab}}=[0,1]$. Such an unbalanced route dataset would result in low classification accuracy. Therefore,
before we randomly shuffle the samples under varying SNRs, we remove extra samples to keep the balance of the route dataset.

{\emph{Remark 2:} We choose the OAMPNet and the IDetNet detectors as two network branches for their different computational complexities and varying  accuracies under different system conditions\footnote{See more details in Tab.~\ref{tabcompass} and Section~\ref{secsumuab}.}. In this case, RouteNet is expected  not only to select a detector with better accuracy under different system conditions, but also to select a detector with lower  complexity without making any compromise on accuracy.
Furthermore, RouteNet can be easily extended to  the route of multiple parallel detectors by  generalizing the binary classification label Eq.~\eqref{equdlabgg} into a multi-classification label. Moreover, conventional detectors  could also serve as branches. By adding one more detector with lower complexity as a branch, the complexity of the whole network would be  reduced.   By adding one more detector with better accuracy  as a branch, the accuracy of the whole network would be improve.
For example, we could build the LMMSE, the IDetNet, the OAMPNet and the SD detectors as four parallel branches. Then, by properly designing the route label and the loss function, we could train a route network to select the detector with the lowest
complexity among those exhibiting the same optimal accuracy. Due to the lack of space, the extension to multiple parallel detectors would be left as future work.}

\emph{Remark 3:} Since various
detectors may have different complexities and exhibit different accuracies, we can balance the accuracy and the complexity  by assigning  route labels according to certain rules. For instance, we can define the route label, $\bm r'_{\textrm{Lab}} $, as
\begin{align}\label{equbinady}
\begin{array}{l}
\bm r'_{\textrm{Lab}} = \left\{
\begin{array}{ll}
$[1,0]$,\quad & {\mathcal{ F}}_{\mathrm{BE}}(\hat{\bm x}_{\textrm{D1}})- {\mathcal{ F}}_{\mathrm{BE}}(\hat{\bm x}_{\textrm{D2}})\leq \epsilon;\\
$[0,1]$, \quad & \textrm{otherwise},
\end{array} \right.
\end{array}
\end{align}
where $\hat{\bm x}_{\textrm{D1}}$ and $\hat{\bm x}_{\textrm{D2}}$ are respectively the estimates of  Detector-1 and   Detector-2. If the complexity of  Detector-1 is higher than  Detector-2, then we can select  a proper $\epsilon<0$ such that RouteNet only uses   Detector-1 when  the accuracy gain of  Detector-1 relative to Detector-2 reaches   a certain value. Otherwise, we set $\epsilon>0$ if the complexity of  Detector-1 is lower than  Detector-2.
The flexibility design of the route label endows the proposed DDNet with great potential in real applications.


\subsection{Centralized Training Steps and Complexity Analysis}\label{seccentr}

\textbf{Centralized training steps:} In CL, the whole training dataset $\mathbb{D}_{\textrm{whole}}$ can be obtained by uploading  the local datasets of all clients to the central server. Then, the training process is executed by the central server. The detail training steps of the CL-DDNet detector are given as follows:
\begin{enumerate}[(a)]
\item Initialize  $\bm \Omega_{\mathrm{ID}}$:    $\{ \beta_k\}_{k=1}^{K_{\textrm{ID}}}=0.7$; $\{\alpha_{k,i}\}_{i,k=1}^{2,K_{\textrm{ID}}}=0.8$; $\{\bm w_{k,i},\bm b_{k,i}\}_{i,k=1}^{3,K_{\textrm{ID}}} \in  \mathcal{N}(\bm{0,0.01I})$.
\item Update  $\bm \Omega_{\mathrm{ID}}$ by using the  ADAM algorithm  to minimize $\mathcal{L}_{\textrm{ID}}\left(\bm \Omega_{\mathrm{ID}}\right)$ until convergence, and then obtain the trained parameter: $\bm \Omega_{\mathrm{ID}}^{*} \leftarrow \bm \Omega_{\mathrm{ID}}$.
\item Initialize  $\bm \Omega_{\mathrm{OA}}$: $\{\gamma_{i,k}\}_{i,k=1}^{3,K_{\textrm{OA}}}=1.0$; $\{\gamma_{4,k}\}_{k=1}^{K_{\textrm{OA}}}=0.0$.
\item Update  $\bm \Omega_{\mathrm{OA}}$ by using the  ADAM algorithm  to minimize $\mathcal{L}_{\textrm{OA}}\left(\bm \Omega_{\mathrm{OA}}\right)$ until convergence, and then obtain the trained parameter: $\bm \Omega_{\mathrm{OA}}^{*} \leftarrow \bm \Omega_{\mathrm{OA}}$.
\item Obtain the  route  dataset $\{(\bm H, \sigma_{n}^{2}),(\bm r_{\textrm{Lab}})\}$ by Eq.~\eqref{equdlabgg}.
\item Initialize $\bm \Omega_{\textrm{RO}}$: $\{\bm w_{\textrm{R},i},\bm b_{\textrm{R},i}\}_{i=1,2} \in  \mathcal{N}(\bm{0,0.01I})$.
\item Update  $\bm \Omega_{\mathrm{RO}}$ by using the  ADAM algorithm  to minimize $\mathcal{L}_{\textrm{RO}}\left(\bm \Omega_{\mathrm{RO}}\right)$ until convergence, and then obtain the trained parameter: $\bm \Omega_{\mathrm{RO}}^{*} \leftarrow \bm \Omega_{\mathrm{RO}}$.
\end{enumerate}
Note that Step (a) $\sim$ (b) and Step (c)  $\sim$ (d) can be executed in parallel.

\begin{table}[!t]\small
\centering
\caption{Complexity comparisons of the detectors}
\label{tabcompass}
\begin{tabular}{|c|c|c|}
\hline
Detector & Computational complexity & {Trainable parameters }  \\
\hline
LMMSE & $O(N_{t}^{3})$ & 0  \\
\hline
OAMPNet & $O(K_{\textrm{OA}}N_{t}^{3})$ & 32 \\
\hline
DetNet & $O(K_{\textrm{ID}}N_{t}^{2})$ & 249600  \\
\hline
IDetNet & $O(K_{\textrm{ID}}N_{t}^{2}+K_{\textrm{ID}}N_{t})$ & 249720   \\
\hline
DDNet & {$O(K_{\textrm{ID}}N_{t}^{2}+K_{\textrm{ID}}N_{t}+N_{t}^{2}) \sim O( K_{\textrm{OA}}N_{t}^{3})$ } &389401 \\
\hline
\end{tabular}
\end{table}
\begin{table}[!t]\small
\centering
\caption{Default Parameters of DDNet}
\label{tabse3}
\begin{tabular}{|c|c|c|c|c|}
\hline
Layer &
$\{{\mathcal{ F}}_{\mathrm{Den}}^{k,1}\}_{k=1}^{K_{\textrm{ID}}}$ &
$\{{\mathcal{ F}}_{\mathrm{Den}}^{k,i}\}_{k,i=1,2}^{K_{\textrm{ID}},3}$ &
$\mathcal{F}_{\mathrm{Den}}^{\textrm{R},1}$ &
$\mathcal{F}_{\mathrm{Den}}^{\textrm{R},2}$\\
\hline
Neurons & 64 & 32 & 128 & 2 \\
\hline
\end{tabular}
\end{table}

\textbf{Complexity analysis: }
 The computational complexities of the LMMSE, the OAMPNet, the DetNet, the IDetNet and the DDNet detectors are given in Tab.~\ref{tabcompass}.  The LMMSE detector has a complexity of
$O(N_{t}^{3})$ due to the matrix inversion, but requires no training and iteration.
 The OAMPNet detector requires the matrix inversion operation in each layer as shown in  Eq.~\eqref{equbsd}, and therefore has a  complexity of $O(K_{\textrm{OA}}N_{t}^{3})$. The IDetNet detector has slightly higher order of complexity than the DetNet detector  since the  smoothing functions in each layer  introduce the extra complexity of $O(K_{\textrm{ID}}N_{t})$. Moreover, since RouteNet introduces  extra complexity of $O(N_{t}^{2})$ and only activates one of the  OAMPNet and the IDetNet detectors for every sample\footnote{It should be mentioned that in real systems, the inputs of RouteNet, i.e., $\bm H$ and $\sigma_{n}^{2}$, are same for the samples
within coherent time, which implies that only  once prediction of RouteNet is require during the same coherent time, and thus  the computational cost would be further reduced.}, the computational complexity of  the DDNet  detector varies for different samples and
ranges from $O(K_{\textrm{ID}}N_{t}^{2}+K_{\textrm{ID}}N_{t}+N_{t}^{2}) \sim O( K_{\textrm{OA}}N_{t}^{3})$.

Using default parameters of the DDNet  detector in Tab.~\ref{tabse3} as an example,  we compare the numbers of trainable parameters of the LMMSE, the OAMPNet, the DetNet, the IDetNet and the DDNet detectors  in Tab.~\ref{tabcompass}, where
the layer numbers   $K_{\textrm{ID}}$ and $K_{\textrm{OA}}$ are set to be 40 and 8, respectively\footnote{The specifical values of these  default parameters are basically selected by trails and errors such that these algorithms perform well.}.
As shown in Tab.~\ref{tabcompass}, the OAMPNet  detector has much fewer    trainable parameters but  higher computational complexity than the IDetNet detector.


\begin{figure}[!t]
\centering
\includegraphics[width=100 mm]{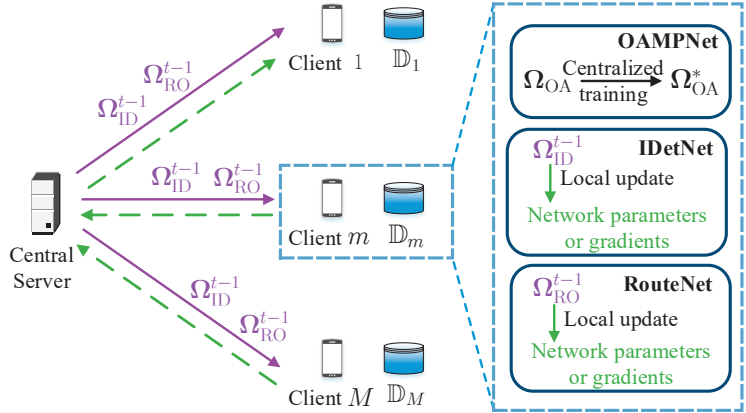}
\caption{Federated training process of the DDNet detector, where ${\mathbb{D}}_{m}$ is the local dataset at the $m$-th client. {Note that the number of selected clients $M$  does not have to be consistent  during  the federated training processes of IDetNet and RouteNet. }}
\label{figfd}
\end{figure}
\section{Federated Dynamic Detection Network}\label{secfddse}
FL protects the privacy of clients by leaving the raw data to local clients and only uploading model parameters to the central server for aggregation.
Since OAMPNet requires very few training samples and training epoches, we will only  train OAMPNet on local clients  in the CL way\footnote{ Based on our experiments, only 100 training samples and a single training epoch
are sufficient to make OAMPNet converge without overfitting.}, while train IDetNet and RouteNet  in FL  way.
 Fig.~\ref{figfd} illustrates the federated training process of DDNet, where $M$ clients are randomly selected  to perform local updates during each global epoch.
Let   $\bm \Omega_{\mathrm{ID}}^{t}$ and $\bm \Omega_{\mathrm{RO}}^{t}$ respectively represent the global network parameters of IDetNet and RouteNet  at the end of  the $t$-th  global epoch.
The federated training process of IDetNet/RouteNet includes  four steps:
\begin{enumerate}[(i)]
\item \textbf{Weights Broadcast:} The central server broadcasts the global network parameter ($\bm \Omega_{\mathrm{ID}}^{t-1}$ or $\bm \Omega_{\mathrm{RO}}^{t-1}$) to each local client. 
\item \textbf{Local update:} Each selected client updates their local network parameters or only calculates the local gradients of the network parameters on their local  datasets  in parallel.
\item \textbf{Parameter upload:} Each selected client uploads their  local network parameters or the local gradients  to the central server.
\item \textbf{Global aggregation:} The central server obtains  global network parameters  ($\bm \Omega_{\mathrm{ID}}^{t}$ or $\bm \Omega_{\mathrm{RO}}^{t}$)  based on the parameters received from the clients.
\end{enumerate}

In the following, we will  present the  FedAve-DDNet detector in detail. To further reduce the transmission overhead, we will also  propose the FedGS-DDNet detector.
\subsection{The FedAve-DDNet Detector}
Denote the detection samples available at the $m$-th client  as ${\mathbb{D}}_{m}^{\textrm{D}}$, with
${\mathbb{D}}_{m}^{\textrm{D}}\buildrel \Delta \over=\{(\bm y, \bm H, \sigma_{n}^{2}),(\bm x)\}$ and $|{\mathbb{D}}_{m}^{\textrm{D}}|={D}_{m}^{\textrm{D}}$.
We first train IDetNet following the above Step (i) $\sim$ (iv). In Step (ii), each selected client updates their local network parameters by   $\ell_m$ successive parameter updates. During the $t$-th global epoch, the updated local network parameter of the $m$-th client,  denoted as $\bm \Omega_{\mathrm{ID}}^{t,m}$,   can be obtained  by
\begin{align}
\bm \Omega_{\mathrm{ID}}^{t,m}  \leftarrow   \textrm{ADAM}^{\ell_m}(\bm \Omega_{\mathrm{ID}}^{t-1}, \nabla_{\bm \Omega_{\mathrm{ID}}}^{{\mathbb{D}}_{m}^{\textrm{D}}}\mathcal{L}_{\textrm{ID}} ),
\end{align}
where $\nabla_{\bm \Omega_{\mathrm{ID}}}^{{\mathbb{D}}_{m}^{\textrm{D}}}\mathcal{L}_{\textrm{ID}}$ represents the gradients of  $\mathcal{L}_{\textrm{ID}}$ with respect to the network parameter $\bm \Omega_{\mathrm{ID}}$ on the dataset ${\mathbb{D}}_{m}^{\textrm{D}}$. Moreover,
 $\textrm{ADAM}^{\ell_m}(\bm \Omega_{\mathrm{ID}}^{t-1}, \nabla_{\bm \Omega_{\mathrm{ID}}}^{{\mathbb{D}}_{m}^{\textrm{D}}}\mathcal{L}_{\textrm{ID}})$
represents  $\ell_m$ successive parameter-updates
by  the ADAM algorithm \cite{kingmaadam}, where $\bm \Omega_{\mathrm{ID}}^{t-1}$ is the initial network parameter of the first parameter-update. In Step (iii), each selected client uploads their updated local parameters, i.e.,
$\{\bm \Omega_{\mathrm{ID}}^{t,m}\}_{m\in \mathbb{M}_{\textrm{ID}}}$, to the central server, where $\mathbb{M}_{\textrm{ID}}$ represents  the set of selected clients during  the federated training processes of IDetNet.
In Step (iv),
the central server obtains   global network parameters by  weighted aggregation.  The global network parameter $\bm \Omega_{\mathrm{ID}}^{t}$ is given by
\begin{eqnarray}\label{equagid}
\bm \Omega_{\mathrm{ID}}^{t} \leftarrow \frac{\sum_{m\in \mathbb{M}_{\textrm{ID}}}{D}_{m}^{\textrm{D}}\bm \Omega_{\mathrm{ID}}^{t,m}}{\sum_{m\in \mathbb{M}_{\textrm{ID}}}{D}_{m}^{\textrm{D}}}.
\end{eqnarray}

After the training of OAMPNet and IDetNet,  each client can obtain the route  dataset $ {\mathbb{D}}_{m}^{\textrm{R}}\buildrel \Delta \over=\{(\bm H, \sigma_{n}^{2}),(\bm r_{\textrm{Lab}})\} $ for RouteNet by Eq.~\eqref{equdlabgg}.
Then, we will train  RouteNet in the similar way with  IDetNet.  More specifically, the weight updates in the clients and the central server can be respectively given as follows:
\begin{eqnarray}\label{equsfgf}
\bm \Omega_{\mathrm{RO}}^{t,m} \!\! &\leftarrow  & \!\! \textrm{ADAM}^{\ell_m}(\bm \Omega_{\mathrm{RO}}^{t-1},\nabla_{\bm \Omega_{\mathrm{RO}}}^{{\mathbb{D}}_{m}^{\textrm{R}}}\mathcal{L}_{\textrm{RO}}), \\
\bm \Omega_{\mathrm{RO}}^{t} \!\!& \leftarrow & \!\! \frac{\sum_{m\in \mathbb{M}_{\textrm{RO}}}{D}_{m}^{\textrm{R}}\bm \Omega_{\mathrm{RO}}^{t,m}}{\sum_{m\in \mathbb{M}_{\textrm{RO}}}{D}_{m}^{\textrm{R}}},
\end{eqnarray}
where ${D}_{m}^{\textrm{R}}=|{\mathbb{D}}_{m}^{\textrm{R}}|$,  $\mathbb{M}_{\textrm{RO}}$ represents  the set of selected clients during  the federated training processes of RouteNet, and $\bm \Omega_{\mathrm{RO}}^{t,m}$ is the local network parameter of RouteNet for the $m$-th client during the $t$-th  global epoch.

The concrete  training steps  of the FedAve-DDNet detector  are given in \textbf{Algorithm~\ref{notr}}, where $T_{\mathrm{F,ID}}$ and $T_{\mathrm{F,RO}}$ are respectively the numbers of  global epoches required for the training of IDetNet and RouteNet.
Note that the training of OAMPNet, the function \emph{LocalUpdateUpload},  and the function  \emph{RouteDataset} are all executed by local clients.
Moreover, the training of OAMPNet and the local updates of IDetNet can be executed in
parallel.

\begin{algorithm}[!t]
\caption{The FedAve-DDNet detector}
\label{notr}
\textbf{Central server executes:}
\begin{algorithmic}[1]
\State {/*Train IDetNet*/}
\State Initialize  $\bm \Omega_{\mathrm{ID}}$  following Step (a)   in Section~\ref{seccentr}
\For{$t=1,2,\cdots,T_{\mathrm{F,ID}}$}
\State Broadcast $\bm \Omega_{\mathrm{ID}}^{t-1}$ to randomly selected $M$ clients
\For{each client $m=1$ to $M_{\mathrm{ID}}$ \textbf{in parallel}}
 \State { \emph{LocalUpdateUpload}}{($m,\bm \Omega_{\mathrm{ID}}^{t-1}; \mathcal{L}_{\mathrm{ID}}, \ell_m,{\mathbb{D}}_{m}^{\textrm{D}}$)}
\EndFor
\State $\bm \Omega_{\mathrm{ID}}^{t} \leftarrow $ \emph{Aggregation}($\{\bm \Omega_{\mathrm{ID}}^{t,m}\}_{m\in \mathbb{M}_{\mathrm{ID}}}, \{D_{m}^{\textrm{D}}\}_{m\in \mathbb{M}_{\mathrm{ID}}}$)
\EndFor
\State Broadcast $\bm \Omega_{\mathrm{ID}}^{T_{\mathrm{F,ID}}}$ to all clients
\State {/* Train RouteNet*/}
\State Initialize $\bm \Omega_{\mathrm{RO}}$ following Step (f)  in Section~\ref{seccentr}
\State ${\mathbb{D}}_{m}^{\textrm{R}} \leftarrow$ \emph{RouteDataset}{($m,\bm \Omega_{\mathrm{ID}}^{T_{\mathrm{F,ID}}},\bm \Omega_{\mathrm{OA}}^*$)}
\For{$t=1,2,\cdots, T_{\mathrm{F,RO}}$}
\State Broadcast $\bm \Omega_{\mathrm{RO}}^{t-1}$ to randomly selected $M$ clients
\For{each client $m=1$ to $M_{\mathrm{RO}}$ \textbf{in parallel}}
 \State { \emph{LocalUpdateUpload}}{($m,\bm \Omega_{\mathrm{RO}}^{t-1}; \mathcal{L}_{\mathrm{RO}}, \ell_m,  {\mathbb{D}}_{m}^{\textrm{R}}$)}
\EndFor
\State $\bm \Omega_{\mathrm{RO}}^{t} \leftarrow $ \emph{Aggregation}($\{\bm \Omega_{\mathrm{RO}}^{t,m}\}_{m\in \mathbb{M}_{\mathrm{RO}}}, \{D_{m}^{\textrm{R}}\}_{m\in \mathbb{M}_{\mathrm{RO}}}$)
\EndFor
\State Broadcast $\bm \Omega_{\mathrm{RO}}^{T_{\mathrm{F,RO}}}$ to all clients
\end{algorithmic}
\begin{algorithmic}
\State {\textbf{def} \emph{Aggregation}}{($\{\bm \Omega^{t,m}\}_{m\in \mathbb{M}}, \{D_{m}\}_{m\in \mathbb{M}}$)}
  \State{\quad$
\bm \Omega^{t} \leftarrow  {\sum_{m\in \mathbb{M}}{D}_m\bm \Omega^{t,m}}/{\sum_{m\in \mathbb{M}}{D}_m}$}
\State\quad {\Return $\bm \Omega^{t}$}
\end{algorithmic}
\textbf{Clients execute:}
\begin{algorithmic}
\State {Train OAMPNet following Step (c) $\sim$ (d) in Section~\ref{seccentr}}
\State {\textbf{def} \emph{LocalUpdateUpload}}{($m,\bm \Omega^{t-1}, \mathcal{L}, \ell_m,\mathbb{D}_m$)}
  \State{\quad$
\bm \Omega^{t,m} \leftarrow \textrm{ADAM}^{\ell_m}(\bm \Omega^{t-1},\nabla_{\bm \Omega}^{\mathbb{D}_m}\mathcal{L})$}
\State {\quad Upload  $\bm \Omega^{t,m} $ to central server}
\State {\textbf{def} \emph{RouteDataset}}{($m,\bm \Omega_{\mathrm{ID}},\bm \Omega_{\mathrm{OA}}$)}
\State \quad Obtain the dataset ${\mathbb{D}}_{m}^{\textrm{R}}$ with Eq.~\eqref{equdlabgg}
\State\quad {\Return ${\mathbb{D}}_{m}^{\textrm{R}}$}
\end{algorithmic}
\end{algorithm}

\subsection{The FedGS-DDNet Detector}
Different from the FedAve-DDNet detector where the clients upload local network parameters ($\bm \Omega_{\mathrm{ID}}^{t,m}$ or $\bm \Omega_{\mathrm{RO}}^{t,m}$) to  the central server, the FedGS-DDNet detector requires the clients to upload  locally sparsified   gradients, denoted by $\mathcal{S}(\bm g_{\mathrm{ID}}^{t,m})$ or $\mathcal{S}(\bm g_{\mathrm{RO}}^{t,m})$, to the central server. Next, the central server would  update the global network parameters by one time ADAM optimization, which can be written as following:
 \begin{align}
\bm \Omega_{\mathrm{ID}}^{t} \leftarrow \textrm{ADAM}^{1}(\bm \Omega_{\mathrm{ID}}^{t-1},\bm g_{\mathrm{ID}}^{t}),\\
\bm \Omega_{\mathrm{RO}}^{t} \leftarrow \textrm{ADAM}^{1}(\bm \Omega_{\mathrm{RO}}^{t-1},\bm g_{\mathrm{RO}}^{t}),
\end{align}
where $\bm g_{\mathrm{ID}}^{t}$ and $\bm g_{\mathrm{RO}}^{t} $ are weighted gradients   given by
 \begin{align}
\bm g_{\mathrm{ID}}^{t} \leftarrow \frac{\sum_{m\in \mathbb{M}_{\textrm{ID}}}{D}_{m}^{\textrm{D}}\mathcal{S}(\bm g_{\mathrm{ID}}^{t,m})}{\sum_{m\in \mathbb{M}_{\textrm{ID}}}{D}_{m}^{\textrm{D}}}, \\
\bm g_{\mathrm{RO}}^{t} \leftarrow \frac{\sum_{m\in \mathbb{M}_{\textrm{RO}}}{D}_{m}^{\textrm{R}}\mathcal{S}(\bm g_{\mathrm{RO}}^{t,m})}{\sum_{m\in \mathbb{M}_{\textrm{RO}}}{D}_{m}^{\textrm{R}}}.
\end{align}

The core problem for  the FedGS-DDNet  is how to sparsify the local gradients, i.e., how to obtain $\mathcal{S}(\bm g_{\mathrm{ID}}^{t,m})$ or $\mathcal{S}(\bm g_{\mathrm{RO}}^{t,m})$.
The idea of the gradient sparsification technique is to randomly discard some elements of the gradients to reduce the parameter transmission overhead and amplify the rest elements to retain the unbiasedness of  the sparsified  gradients.
More specifically, we first flatten the  gradients of the  network parameter as the gradient vector\footnote{Here the superscript $m$, the  subscript $\textrm{ID}$ and the  subscript  $\textrm{RO}$ are all omitted for simplicity.} $\bm g^t=[g_{1}, \cdots,g_{Q}]$, where $Q$ is the size of the network parameters.
Let $\mu_q\in \{0,1\}$ ($1\leq q\leq Q$) be a binary-valued random variable indicating whether $g_{q}$ is selected. Define the probability of $\mu_q=1$ as  $p_q$, which implies  $g_{q}$ is selected to be uploaded to  the central server with  the probability  $p_q$. Then, the sparsified  gradient $\mathcal{S}(\bm g^t)$  can be written as
\begin{equation}\label{equggss}
\mathcal{S}(\bm g^{t}) =[ \frac{\mu_1 g_{1}}{p_1}, \cdots,\frac{\mu_Q g_{Q}}{p_Q} ].
\end{equation}
Note that the  sparsified  gradient vector is divided by the probability vector $\bm p=[p_1, \cdots,p_Q]$  in element-wise manner to retain the unbiasedness of  the sparsified  gradients, i.e., $E[\mathcal{S}(\bm g^{t})]=E[\bm g^{t}]$.
The  sparsity parameter  $\delta$ is defined as  
\begin{equation}\label{equqq}
\delta=\sum\nolimits_{q=1}^{Q} \frac{p_q}{Q}.
\end{equation}
 To reduce the parameter transmission overhead, we expect $\delta$  to be as small as possible. However, a small  $\delta$  would also degrade the   performance of IDetNet  or RouteNet. Besides, the specific values of the  probability vector $\bm p$ should also be taken into account for the sake of better performance.

 To achieve a better tradeoff between the accuracy and the sparsity, we
consider a simplified optimization problem, where $\mathcal{L}$ and $\bm \Omega^{t}$   respectively denote the loss function and  the network parameter in the $t$-th iteration.
The gradient $\bm g^{t}$ is estimated based on  a random batch of data samples, and   is an unbiased  estimate of the true gradient $\nabla_{\bm \Omega^{t}} \mathcal{L}$, i.e.,
  $ E [ \bm g^{t}]=\nabla_{\bm \Omega^{t}} \mathcal{L}$.
Assume the loss function $\mathcal{L}$  satisfies the  Lipschitz continuous gradient condition,  then there exists a constant $L$ such that \cite{nesterov1998introductory}
 \begin{align}
\mathcal{L}(\bm \Omega^{t+1})  \leq \mathcal{L}(\bm \Omega^{t})+\nabla_{\bm \Omega^t} \mathcal{L}^{T}(\bm \Omega^{t+1}\!-\!\bm\Omega^{t})+\frac{L}{2}\|\bm \Omega^{t+1}\!-\!\bm \Omega^{t}\|^{2}   = \mathcal{L}(\bm \Omega^{t})-\eta\nabla_{\bm \Omega^t} \mathcal{L}^{T}\bm g^t+\frac{L}{2}\eta^2\|\bm g^{t}\|^{2} \nonumber
\end{align}
holds, where $\eta$ is the learning rate. Then, we have
 \begin{eqnarray}
E[\mathcal{L}(\bm \Omega^{t+1})]\leq \mathcal{L}(\bm \Omega^{t})-\eta\|\nabla_{\bm \Omega^t} \mathcal{L}\|^2  + \frac{L}{2}\eta^2 E[ \|\bm g^{t}\|^{2}],
\end{eqnarray}
which indicates that the variance  $ E[\|\bm g^{t}\|^{2}]$ has negative impacts on the convergence accuracy, and therefore we should control the  variance of the sparsified  gradients under a certain  level to reduce the negative impacts.
To find an optimal probability vector $\bm p$  that can not only satisfies the variance constraint but also minimizes the sparsity,  we model the tradeoff between the variance and sparsity as an optimization problem  \cite{Wangni10}:
 \begin{align}\label{equaas}
\mathop {\min }\limits_{\bm p} \sum\nolimits_{q=1}^{Q} p_q \quad
 \textrm{s.t.}\  E[|\mathcal{S}^{2}(\bm g^{t})|_1]=\sum\nolimits_{q=1}^{Q} \frac{g_{i}^{2}}{p_i} \le (1+\varsigma)\sum\nolimits_{q=1}^{Q}  {g_{i}^{2}},
\end{align}
where $\varsigma$ is the parameter to limit the variance of the sparsified  gradient  $\mathcal{S}(\bm g^t)$. Since Eq.~\eqref{equaas} is a classical convex optimization problem, we can   obtain the solution ${p_q}=\min\{\lambda |g_q|,1\}$ by using the Karush-Kuhn-Tucker (KKT) condition, where $\lambda>0$ is an unknown constant. The  solution implies that the   gradients with larger absolute value should be selected with higher  probability.
 Although the value of $\lambda$ can be calculated in closed-form  by the water-filling algorithm (see more details in \cite{Wangni10}),  we only focus on a computational efficient method to approximately solve the problem. More specifically, by initializing  $p_q\leftarrow\min\{\delta Q |g_q|/\sum_{q} |g_q|,1 \}$, we can obtain the set $\mathbb{Q}\leftarrow \{1\leq q\leq Q|p_q<1\}$ and calculate the  amplification coefficient $a$ for the $\{p_q\}_{ q\in \mathbb{Q}}$   by meeting Eq.~\eqref{equqq}, i.e.,
 \begin{equation}\label{equqqe}
 a\sum\nolimits_{q\in \mathbb{Q}} |p_q|+Q-|\mathbb{Q}|=\delta Q.
\end{equation}
By iteratively amplifying  $\{p_q\}_{ q\in \mathbb{Q}}$ with $p_q\leftarrow\min\{ap_q,1\}$, updating the set $\mathbb{Q}$, and calculating $a$ with Eq.~\eqref{equqqe}  until $a$ is close enough to 1, the optimal  probability vector $\bm p$ and the optimal sparsified  gradient $\mathcal{S}(\bm g^t)$  can be obtained. The gradient sparsification technique only involves addition, multiplication, and minimization operations, and therefore is computationally
efficient, especially on parallel computing hardware like graphic processing units (GPUs).

The concrete training steps  of the FedGS-DDNet detector  are the same with  \textbf{Algorithm~\ref{notr}} except for the functions \emph{LocalUpdateUpload} and  \emph{Aggregation} that should be respectively replaced with the functions \emph{SLocalUpdateUpload} and  \emph{SAggregation}, as shown in  \textbf{Algorithm~\ref{notred}}.

\begin{algorithm}[!t]
\caption{Core functions of the FedGS-DDNet detector}
\label{notred}
{\textbf{def} \emph{SAggregation}}{($\{\mathcal{S}(\bm g^{t,m})\}_{m\in \mathbb{M}}, \{D_{m}\}_{m\in \mathbb{M}}$)}
\begin{algorithmic}
  \State{$
\bm g^{t} \leftarrow  {\sum_{m\in \mathbb{M}}{D}_m\mathcal{S}(\bm g^{t,m})}/{\sum_{m\in \mathbb{M}}{D}_m}$}
\State{$\bm \Omega^{t} \leftarrow \textrm{ADAM}^{1}(\bm \Omega^{t-1},\bm g^{t})$}
\State {\Return $\bm \Omega^{t}$}
\end{algorithmic}
{\textbf{def} \emph{SLocalUpdateUpload}($m,\bm \Omega^{t-1}, \mathcal{L}, \mathbb{D}_m$)}
\begin{algorithmic}
\State{ $\bm g^{t,m} \leftarrow \textrm{Flatten} \nabla_{\bm \Omega^{t-1}}^{\mathbb{D}_m}\mathcal{L}$}
\State{ $p_q\leftarrow\min\{\delta Q |g_{q}^{t,m}|/\sum_{q} |g_{q}^{t,m}|,1 \}$, $1\leq q\leq Q$}
\While{$c>1+10^{-2}$}
\State{ $\mathbb{Q}\leftarrow\{1\leq q\leq Q|p_q<1\}$}
\State{ $a\leftarrow(\delta Q-Q+|\mathbb{Q}|)/\sum\nolimits_{q\in \mathbb{Q}} |p_q|$}
\State{ $p_q\leftarrow\min\{ap_q,1\}$, $\{p_q\}_{ q\in \mathbb{Q}}$}
\EndWhile
\State{Obtain the sparsified  gradient $\mathcal{S}(\bm g^{t,m})$ using Eq.~\eqref{equggss} }
\State {Upload   $\mathcal{S}(\bm g^{t,m})$  to central server}
\end{algorithmic}
\end{algorithm}

\subsection{Transmission Overhead}
The transmission overhead of the CL-DDNet detector includes the transmission of the whole training dataset $\mathbb{D}_{\textrm{whole}}$.
Denote $q^{(d)}$ as the parameter size of the $d$-th sample $\{(\bm y, \bm H, \sigma_{n}^{2}),(\bm x)\}$, which can be calculated as
$q^{(d)}=2N_{r}^{(d)}+4N_tN_{r}^{(d)}+2N_t+1$. Then, the transmission overhead of the CL-DDNet detector   can be written as
\begin{eqnarray}\label{equtradnse}
\mathcal{T}_{\textrm{CL}} = b\sum\nolimits_{d=1}^{|\mathbb{D}_{\textrm{whole}}|} q^{(d)},
\end{eqnarray}
where $b$ is the number of bits required to represent a floating-point number.
In contrast, the transmission overhead of the  FedAve-DDNet detector  includes the
transmission of the network parameters, i.e., $\{\bm \Omega_{\mathrm{RO}}^{t,m},\bm \Omega_{\mathrm{RO}}^{t},\bm \Omega_{\mathrm{ID}}^{t,m},\bm \Omega_{\mathrm{ID}}^{t}\}$,  during the weights broadcast and the parameter upload processes.
 Hence, the transmission overhead of the  FedAve-DDNet detector is given by
 \begin{eqnarray}\label{equtsrrdd}
\mathcal{T}_{\textrm{FedAve}} = 2bQ_{\mathrm{ID}}T_{\mathrm{F,ID}}M_{\mathrm{ID}}+2bQ_{\mathrm{RO}}T_{\mathrm{F,RO}}M_{\mathrm{RO}},
\end{eqnarray}
where
 $Q_{\mathrm{ID}}$ and  $Q_{\mathrm{RO}}$ are respectively the trainable-parameter-sizes of IDetNet and  RouteNet.
Furthermore, the transmission overhead of the FedGS-DDNet detector is given by
\begin{align}
\label{Eqqccon}
\mathcal{T}_{\textrm{FedGS}}=& \underbrace {b\delta(Q_{\mathrm{ID}}T_{\mathrm{F,ID}}M_{\mathrm{ID}}+Q_{\mathrm{RO}}T_{\mathrm{F,RO}}M_{\mathrm{RO}}) }_{\textrm{gradient upload}}+\underbrace {Q_{\mathrm{ID}}T_{\mathrm{F,ID}}+Q_{\mathrm{RO}}T_{\mathrm{F,RO}}}_{\textrm{index upload}} \nonumber \\
& + \underbrace {b(Q_{\mathrm{ID}}T_{\mathrm{F,ID}}M_{\mathrm{ID}}+Q_{\mathrm{RO}}T_{\mathrm{F,RO}}M_{\mathrm{RO}}) }_{\textrm{weights broadcast}}.
\end{align}
Note that we should upload the index vector, $\bm \mu=[\mu_1,\cdots,\mu_Q]$, during every   global epoch to indicate the index of the nonzero sparsified gradients.
Since the index vector $\bm \mu$ has the length of  $Q_{\mathrm{ID}}$ during the training of IDetNet, it requires $Q_{\mathrm{ID}}T_{\mathrm{F,ID}}$ bits to represent. Similarly, the   transmission overhead during the training of RouteNet is  $Q_{\mathrm{RO}}T_{\mathrm{F,RO}}$.


\section{Simulation Results}\label{secnumer}
In this section, we will  present the implementation details of the proposed detectors,   including default system  and algorithm    parameters. Then, we will investigate the performance of the CL-DDNet, the FedAve-DDNet, and the FedGS-DDNet detectors, followed by the transmission overhead comparisons.

\subsection{Implementation Details}
The data samples are generated by transmitting random QPSK sequences though correlated Rayleigh MIMO channel that can be described by the Kronecker model \cite{9360873} $\tilde{\bm{H}}=\sqrt{\bm{R}_{t}}\tilde{\bm{H}}_g \sqrt{\bm{R}_{r}},$
where $\tilde{\bm{H}}_g$ is the i.i.d. channel with each element following
$\mathcal{N}_{C}({0,1/N_r})$. Moreover, $\bm{R}_{t}$ and $\bm{R}_{r}$ respectively represent the transmitter and the receiver channel correlation matrices, and are given by
\begin{eqnarray*}
\bm{R}_{{t}}=\left[\!\begin{array}{ccc}
1 & \rho & \ldots \\
\rho & 1 & \ldots \\
\vdots & & \ddots \\
\rho^{\left(N_{t}-1\right)^{2}} & & \!\ldots
\end{array}\!\right], \bm{R}_{{r}}=\left[\!\begin{array}{ccc}
1 & \rho & \ldots \\
\rho & 1 & \ldots \\
\vdots & & \ddots \\
\rho^{\left(N_{r}-1\right)^{2}} & &\! \ldots
\end{array}\!\right],
\end{eqnarray*}
where $\rho$ is the correlation coefficient.
In the training stage of FL, $N_t$ is fixed to be 16, while  $N_r$ for each client  is a random variable that uniformly distributes over the interval $[16,64]$.  To generate the local dataset for  a certain client,    $\rho$ is a random subinterval of length 0.2 in the  interval [0,0.9], and SNR is a random subinterval of length 5 dB in the  interval [-5,15] dB.
 In the training stage of CL,   the whole dataset ${\mathbb{D}}_{\textrm{whole}}$  is obtained by collecting and randomly shuffling the local datasets of all the local datasets in clients.
In the testing stage of FL, the bit-error-rate  (BER) performance is evaluated  over the  testing samples that are generated following the same way as  ${\mathbb{D}}_{\textrm{whole}}$.
In the testing stage of CL, the BER performance is evaluated  over the  testing samples that are generated under specified SNR, transmitted antenna number, and correlation coefficient conditions.

All the networks are   implemented on the same computer with one Nvidia GeForce GTX 1080 Ti GPU.  TensorFlow 2.4  is employed as the DL framework.
 The penalty coefficient $\xi$ is 0.5.
The initial learning rate of the ADAM optimizer is 0.001 and decays  by a factor of 0.9 whenever the verification loss does not  drop within 20 epoches.

\subsection{The CL-DDNet Detector}\label{secsumuab}

\begin{figure}[!t]
\centering
\includegraphics[width= 85mm]{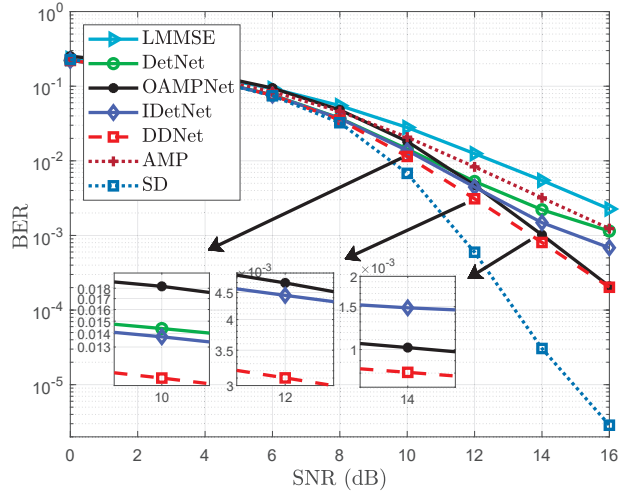}
\caption{{BER performance comparisons of  DDNet with other MIMO detectors, testing at varying SNRs.}}
\label{figdsnr}
\end{figure}

In this subsection, all networks are trained in the CL way.
{Fig.~\ref{figdsnr} compares the BER performance of the LMMSE,  the SD \cite{1194444},  the AMP \cite{6778065}, the DetNet,
the OAMPNet, the IDetNet and the DDNet detectors} at varying  SNRs. The  IDetNet detector  can achieve better BER performance than the DetNet detector, and the performance gain increases as  SNR increases, which validates the effectiveness of  the improvement  skills in Section~\ref{secidetnet}. {Furthermore, the IDetNet detector outperforms all other  detectors except for   the DDNet and the SD detectors  when SNR is lower than 12 dB. The OAMPNet  detector outperforms all other  detectors except for  the DDNet and the SD detectors  when  SNR is higher than  12 dB.
 Moreover, the DDNet detector consistently outperforms all other detectors except for the SD detector at all SNRs,} and its performance gain compared to the upper bound of the IDetNet and  the OAMPNet detectors is more significant when the performance of the IDetNet and the  OAMPNet detectors is  closer, as illustrated in  the three enlarge pictures. This is because  when the performance gap between the IDetNet and  the OAMPNet detectors is wide at a certain SNR, the  subnetwork RouteNet of the DDNet detector basically selects the definitely  better one between  IDetNet and OAMPNet for the testing samples at the certain SNR. Therefore the  DDNet detector only achieves slight performance gains over the better one among the IDetNet and the OAMPNet detectors. In contrast, when the performance gap between the IDetNet and the  OAMPNet detectors is narrow, the   sample-wise dynamic routing among  IDetNet and OAMPNet would bring more significant performance gains to the  DDNet detector.

 \begin{table}[!t]\small
\centering
\caption{ {Average FLOPs of the detectors}}
\label{tabcflp}
\begin{tabular}{|c|c|c|c|c|c|c|}
\hline
Detector & LMMSE & IDetNet & OAMPNet &  DDNet (12 dB) &  DDNet (16 dB)&  DDNet ($0\sim 16$ dB) \\
\hline
 Flops ($1e5$) & 1.61 & 6.00  & 55.66 &  8.1 & 16.3 & 9.2\\
 \hline
\end{tabular}
\end{table}

 { Moreover,  we calculate the average number of floating point operations (FLOPs),   including  multiplications and divisions, required for one sample with various detectors\footnote{In the calculation process, we assume that the Gauss-Jordan elimination algorithm is adopted to compute the matrix inversion.}, as shown in Tab.~\ref{tabcflp}. The average number of FLOPs  required for the DDNet detector increases as SNR  increases because the OAMP subnetwork is more likely to be activated as  SNR  increases.
 As shown in Fig.~\ref{figdsnr} and Tab.~\ref{tabcflp}, the proposed DDNet detector could achieve better performance than the OAMP detector with much lower complexity.
}

\begin{figure}[!t]
\centering
\includegraphics[width=85mm]{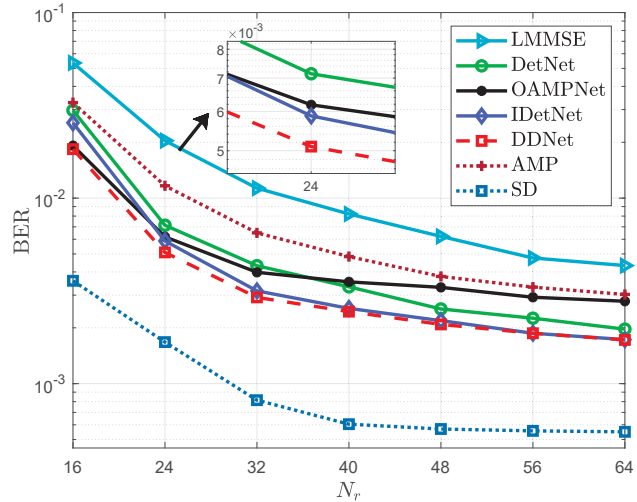}
\caption{{BER performance comparisons of  DDNet with other MIMO detectors, testing at varying values of $N_r$.}}
\label{figfdn}
\end{figure}

{Fig.~\ref{figfdn}  compares the BER performance of  the LMMSE, the SD,  the AMP, the DetNet,
the OAMPNet, the IDetNet and the DDNet detectors} at varying values of  $N_r$ and SNR = 10 dB.
The performance of all the  detectors improves  as $N_r$  increases and
the IDetNet  detector consistently outperforms the DetNet detector   as $N_r$  increases.
{Moreover, the IDetNet detector outperforms all other  detectors except for   the DDNet and the SD detectors  when $N_r$  is larger than  24  while the OAMPNet  detector  outperforms all other  detectors except for  the DDNet and the SD detectors   when   $N_r$  is smaller than about  24.}
Similar with Fig.~\ref{figdsnr}, the DDNet detector  outperforms all other detectors except for the SD detector. Thanks to  the sample-wise dynamic routing  of DDNet, the  performance gain of the DDNet  detector  compared to the upper bound of the IDetNet and  the OAMPNet  detectors is more significant when the performance of the IDetNet and  the OAMPNet detectors is  closer.
\begin{figure}[!t]
\centering
\includegraphics[width=85mm]{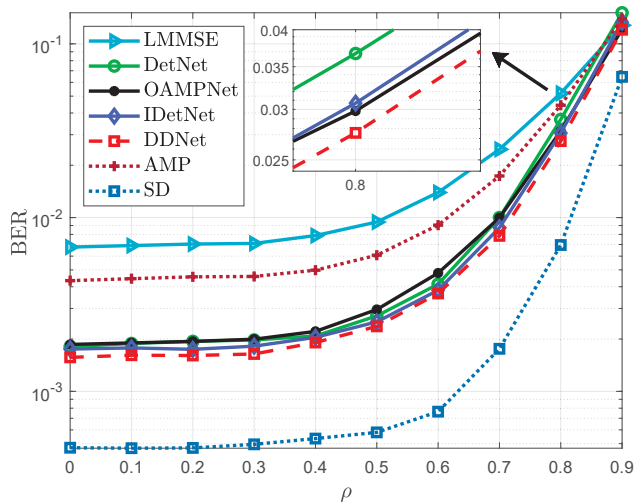}
\caption{{BER performance comparisons of  DDNet with other MIMO detectors, testing at varying correlation coefficient $\rho$.}}
\label{figdrho}
\end{figure}

{Fig.~\ref{figdrho}  compares the BER performance of  the LMMSE, the SD,  the AMP, the DetNet,
the OAMPNet, the IDetNet and the DDNet detectors} at varying values of correlation coefficient  $\rho$ and SNR = 10 dB. The performance of all the  detectors degrades  as $\rho$  increases, and
the IDetNet detector consistently outperforms the DetNet detector as $\rho$ increases.
{ As shown in  the enlarge picture, the IDetNet detector outperforms all other  detectors except for   the DDNet and the SD detectors when $\rho$  is smaller  than    0.8  while the OAMPNet  detector outperforms all other  detectors except for  the DDNet and the SD detectors  when  $\rho$  is larger than   0.8.}
As displayed in Fig.~\ref{figdsnr}~$\sim$~Fig.~\ref{figdrho}, the DDNet detector consistently outperforms all other detectors except for the SD detector under all system conditions, which validates the effectiveness and the superiority of the sample-wise dynamic routing in the DDNet  detector.

\subsection{The FedAve-DDNet Detector}

Tab.~\ref{tabcdcce} investigates  the  impact of varying $M_{\mathrm{ID}}$ and $M_{\mathrm{RO}}$ on the federated training of IDetNet and RouteNet, where the number of local updates $\ell_m$ is set to be 2. We recode the epoch numbers of the IDetNet detector required to reach BER$<10^{-2}$ and the epoch numbers of RouteNet required to  achieve convergence.
It can be seen that a larger number of selected clients leads to  faster convergence but lower computational efficiency. Therefore,
 we respectively set $M_{\mathrm{ID}}$ and $M_{\mathrm{RO}}$  to be 8 and 16 to strike a  balance between the computational efficiency and the convergence rate.

%

\begin{table}[t] \small
  \centering
  \caption{The impact of varying $M_{\mathrm{ID}}$ and $M_{\mathrm{RO}}$ on the training of IDetNet and RouteNet.}
\label{tabcdcce}
    \begin{tabular}{|c|c|cccc||c|c|cccc|}
    \hline
    \multicolumn{2}{|c|}{$M_{\mathrm{ID}}$} & 4 & 8 & 16 & 32 & \multicolumn{2}{c|}{ $M_{\mathrm{RO}}$} & 4 & 8 & 16 & 32  \\
    \hline
    \multirow{2}[2]{*}{\makecell[c]{Epoches\\(BER$<10^{-2}$)} } &   ${D}_{m}^{\textrm{D}}=64$ & 430 & 240 & 180 & 150 & \multirow{2}[2]{*}{\makecell[c]{Epoches\\(Convergence)} } & ${D}_{m}^{\textrm{R}}=64$    & 125 & 80 & 40 & 25 \\
          & ${D}_{m}^{\textrm{D}}=128$   & 380 & 200 & 140 & 110 & & ${D}_{m}^{\textrm{R}}=128$     & 120 & 75 & 40 & 25\\
    \hline
    \end{tabular}%
  \label{tab:addlabel}%
\end{table}%
\begin{figure}[!t]
\centering
\includegraphics[width=85mm]{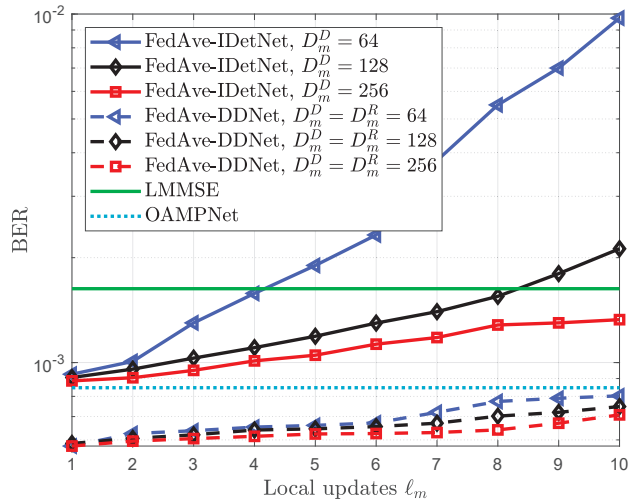}
\caption{BER performance  of the FedAve-DDNet and the FedAve-IDetNet detectors versus the number of  local  updates $\ell_m$.}
\label{figfdstep}
\end{figure}

Fig.~\ref{figfdstep} displays the  BER performance  of the FedAve-DDNet  and  FedAve-IDetNet detectors versus the number of  local  updates $\ell_m$, where the notation ``FedAve-'' is added to indicate that the  detector is trained based on the FedAve algorithm. Note that when $\ell_m$ is 1, the FedAve algorithm is actually equivalent to the CL training with its batch size  equal  to the selected  client number multiplied by the local dataset size.
The performance of the FedAve-DDNet and the FedAve-IDetNet detectors degrades as $\ell_m$ increases while the declining slope decreases as the number of  local samples  ${D}_{m}^{\textrm{D}}$ increases.
This is because that $\ell_m$ successive parameter-updates on the local datasets lead to overfitting, and a smaller  ${D}_{m}^{\textrm{D}}$ results in more serious overfitting.
Moreover,  the LMMSE and the CL based OAMPNet detectors, represented by  two horizontal lines, are used as the benchmarks. It can be seen that the FedAve-IDetNet detector could exhibit worse performance when $\ell_m$ is greater than a certain value.
Furthermore, the FedAve-DDNet  detector consistently outperforms  the OAMPNet detector while their performance  gap becomes narrower  as $\ell_m$ increases. This is because that  the performance gap between the FedAve-IDetNet and OAMPNet detectors becomes wider as $\ell_m$ increases, which indicates that it would be more difficult for  the FedAve-DDNet  detector to benefit from  the sample-wise dynamic routing among the FedAve-IDetNet and OAMPNet.
 Notice that the varying $\ell_m$ has more moderate impacts on the  FedAve-DDNet  detector than  the  FedAve-IDetNet detector, which is due to the stabilization of the CL based OAMPNet detector. It should be mentioned that although the datasets ${\mathbb{D}}_{m}^{\textrm{D}}$ and ${\mathbb{D}}_{m}^{\textrm{R}}$ may not have the same number of samples in actual  implementations according to the Section~\ref{secroutend},    we set ${\mathbb{D}}_{m}^{\textrm{D}}={\mathbb{D}}_{m}^{\textrm{R}}$ in order to   control variable effects, i.e., to better compare the robustness of the FedAve-DDNet and the  FedAve-IDetNet detectors over $\ell_m$.



\begin{figure}[!t]
\centering
\begin{minipage}[t]{0.48\textwidth}
\centering
\includegraphics[width=80mm]{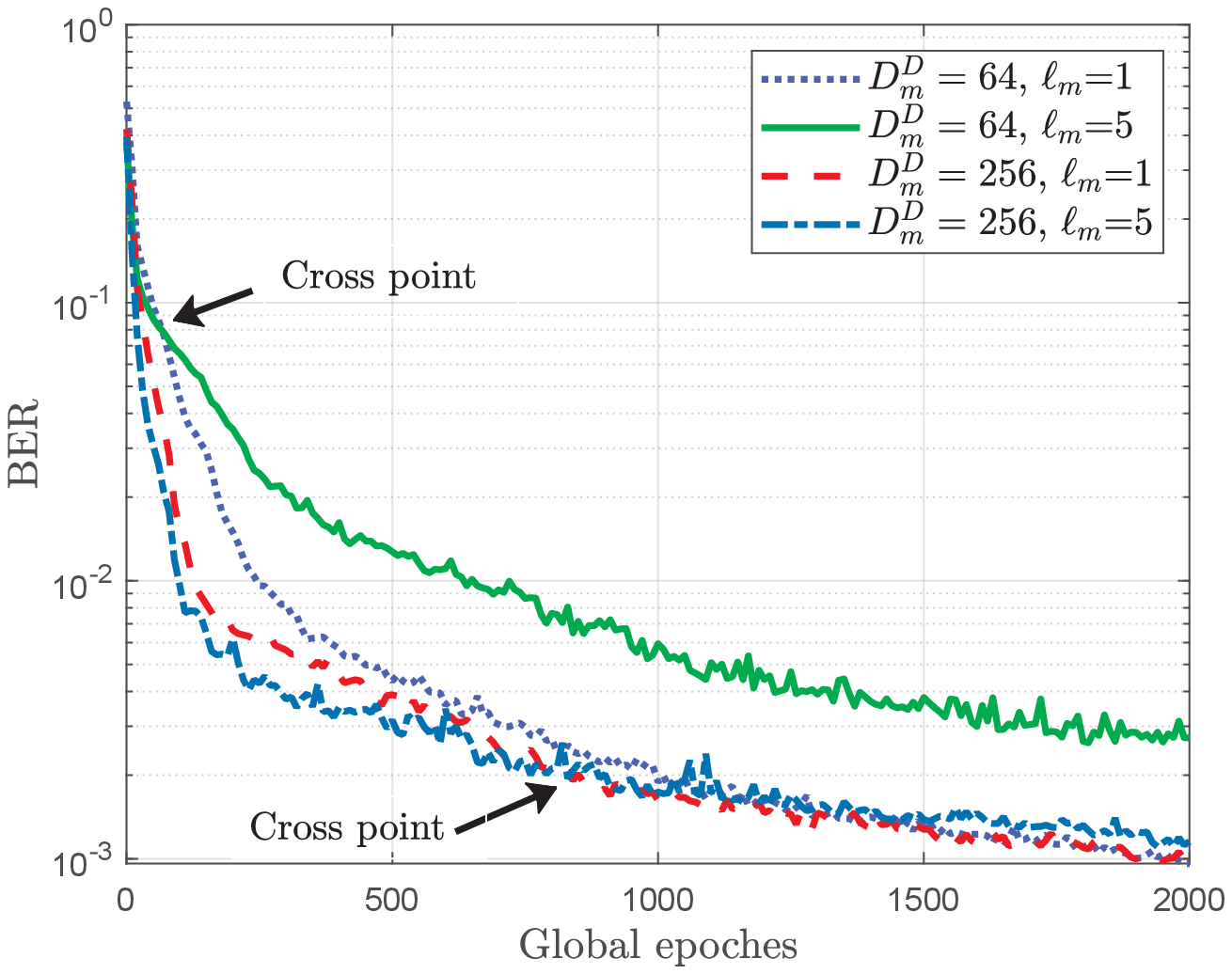}
\caption{BER performance  of  the FedAve-IDetNet detector  versus the number of  global  epoches.}
\label{figfepggr}
\end{minipage}
\begin{minipage}[t]{0.48\textwidth}
\centering
\includegraphics[width=80mm]{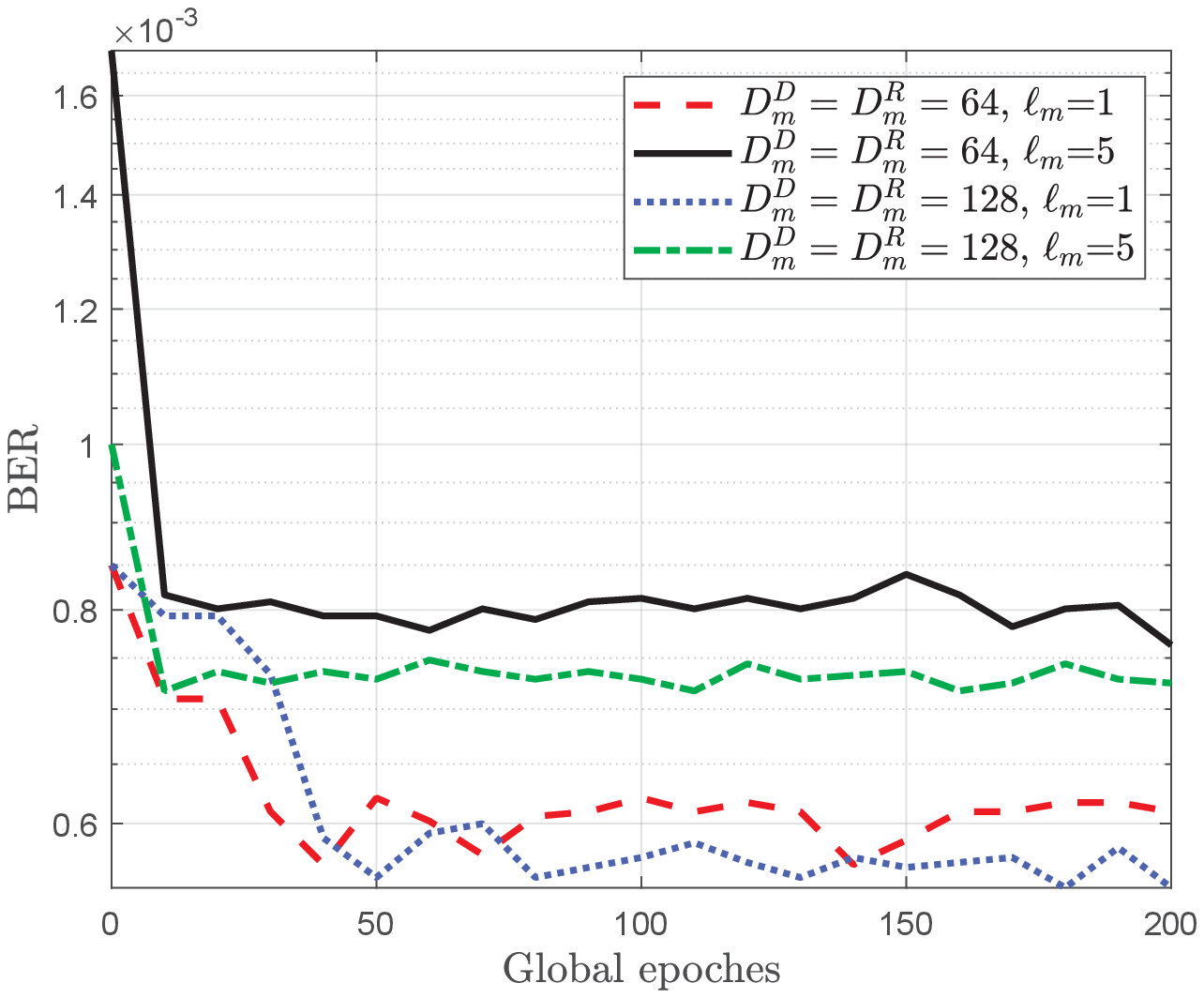}
\caption{BER performance of  the  FedAve-DDNet  detector versus the number of  global  epoches while training  RouteNet.}
\label{figfffep33g}
\end{minipage}
\end{figure}

Fig.~\ref{figfepggr} shows the BER performance  of  the FedAve-IDetNet detectors  versus the number of  global  epoches with varying $\ell_m$ and  ${\mathbb{D}}_{m}^{\textrm{D}}$. The BERs of the FedAve-IDetNet detectors decrease as the number of  global  epoches increases, and the declining slopes also decrease as the number of  global  epoches increases. Notice that the FedAve-IDetNet detectors with  $\ell_m=5$  first decrease  faster   but then decrease  slower   than those with  $\ell_m=1$. The cross point for the  FedAve-IDetNet detectors with  $D_m=64$ occurs during  the 80-th global  epoches while the cross point for the  FedAve-IDetNet detectors with  $D_m=256$ occurs during  the 800-th global  epoches.
Since there is a positive correlation between the transmission overhead  and the number of global epoches as indicated by Eq.~\eqref{equtsrrdd}, we can find that more local updates per global  epoch  could reduce the transmission overhead  in the early stage of training. However, this advantage becomes  insignificant   when  ${\mathbb{D}}_{m}^{\textrm{D}}$ is small,   because that a small local dataset is more likely to be  over-optimized by  local updates, which leads to a
premature  cross point of the BER curves.


Fig.~\ref{figfffep33g} displays the BER performance  of  the FedAve-DDNet  detector  versus the number of  global  epoches in the federated training process of RouteNet with varying $\ell_m$, ${D}_{m}^{\textrm{D}}$ and ${D}_{m}^{\textrm{R}}$. It can be seen that  the FedAve-DDNet  detectors with $\ell_m=1$ converge after training RouteNet 40 global epoches while the FedAve-DDNet  detectors with $\ell_m=5$ converge only after 10 global epoches. Moreover, the value of  ${\mathbb{D}}_{m}^{\textrm{D}}$  or ${D}_{m}^{\textrm{R}}$ almost has no impact on the convergence rate. This is because that RouteNet  has certain tendency of dynamic routing for a certain client, which implies that the datasets of different clients, i.e., $ {\mathbb{D}}_{m}^{\textrm{R}}$, are highly non-i.i.d.. In this case, the route labels of a certain client are
seriously unbalanced, and thus  it would be inefficient to increase the sample number of the local dataset ${\mathbb{D}}_{m}^{\textrm{R}}$ to improve the training performance.

As illustrated in  Fig.~\ref{figfdstep} $\sim$ Fig.~\ref{figfffep33g},  the detectors with larger  $\ell_m$  converge faster in the early stage of training but achieve worse convergence performance.  Therefore, a experienced engineer could fine-tune the value of  $\ell_m$ to improve the converge speed and the final accuracy, i.e., selecting larger  $\ell_m$ at first and decreasing  $\ell_m$ gradually.

\subsection{The FedGS-DDNet Detector}

\begin{figure}[!t]
\centering
\includegraphics[width=85mm]{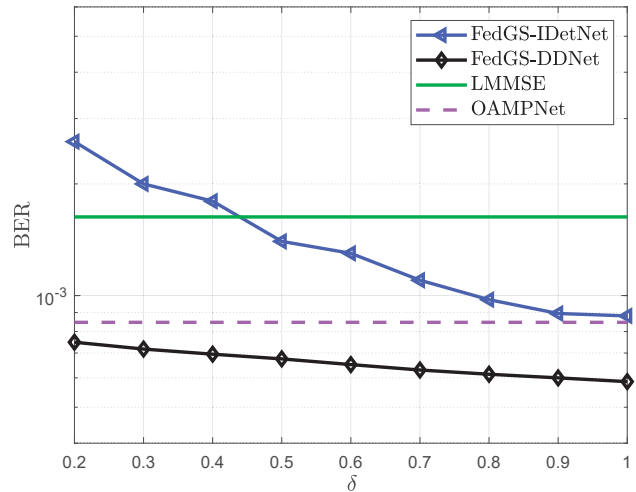}
\caption{BER performance  of the FedGS-DDNet and  FedGS-IDetNet  detectors versus the value of  $\delta$. ${D}_{m}^{\textrm{D}}=256$. ${D}_{m}^{\textrm{R}}=128$.}
\label{figde}
\end{figure}

Fig.~\ref{figde} displays the  BER performance  of the FedGS-DDNet  and  FedGS-IDetNet  detectors versus the value of  $\delta$,  where the notation ``FedGS-'' is added to indicate that the  detector is trained based on the FedGS algorithm.
When $\delta$ is smaller than 0.4,  the  FedGS-IDetNet  detector achieves worse performance than the LMMSE detector. Moreover, the FedGS-DDNet detector consistently outperforms the OAMPNet detector while their performance gap becomes wider as $\delta$ increases, which also implies that the FedGS-DDNet  benefits from the improvements of the  FedGS-IDetNet detector.
 Moreover, the accuracy of both the FedGS-DDNet and  the FedGS-IDetNet detectors decrease  as $\delta$ decreases.
By selecting a proper $\delta$, the proposed FedGS-IDetNet and FedGS-DDNet detectors can strike a good balance between the accuracy and the transmission overhead.

\subsection{Transmission Overhead}
\begin{figure}[!t]
\centering
\includegraphics[width=85mm]{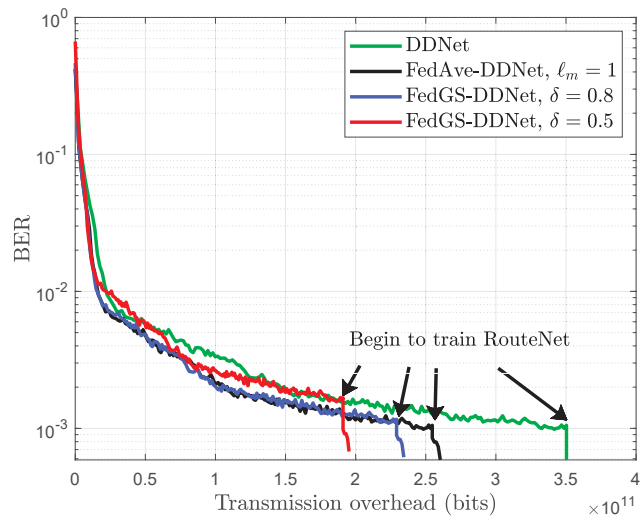}
\caption{ The BER curves of   the CL-DDNet, the FedAve-DDNet,  and the FedGS-DDNet detectors versus the transmission overhead. ${D}_{m}^{\textrm{D}}=256$. ${D}_{m}^{\textrm{R}}=128$. $b=32 $ bits. }
\label{figover}
\end{figure}

Fig.~\ref{figover}  draws the BER curves of  the CL-DDNet, the
FedAve-DDNet, and the FedGS-DDNet detectors   versus the transmission overhead.
Note that we normally begin the training process  of CL  algorithms  after the  whole dataset  is  collected and stuffed at the central server. However,  here we assume that  the dataset collection  and the training are  carried out in parallel during the CL training  in order to intuitively compare the performance gains of the transmission overhead. In other words, Fig.~\ref{figover} can be approximately  interpreted  as the BER performance of the above-mentioned detectors versus the number of global epoches since the transmission overheads of both the datasets and the network parameters are positively correlated with the number of global epoches during the training of IDetNet. Note that the curve of CL-DDNet becomes  vertical when we begin to train the RouteNet since the whole route  dataset  can be generated in the central server based on the trained subnetworks  and the whole dataset ${\mathbb{D}}_{\textrm{whole}}$, thus requiring no additional transmission overhead. As shown in Fig.~\ref{figover}, { the FedAve-DDNet  detector could reduce the transmission overhead  by at least 25.7\%  while maintaining  satisfactory detection accuracy. Moreover, the FedGS-DDNet detector could further reduce the transmission overhead  at the cost of small accuracy loss.}

\section{Conclusion}\label{secconcul}
In this paper, we   designed the architecture of the DDNet detector based on the  sample-wise dynamic routing among   IDetNet and   OAMPNet. By utilizing FL algorithms, we also proposed the FedAve-DDNet and the FedGS-DDNet detectors   to reduce the transmission overhead and protect the  data privacy.
 Simulation results have shown that the proposed DDNet detector could  achieve  better accuracy than both the IDetNet and  the OAMPNet detectors under all system conditions, which validates the superiority  of the  sample-wise dynamic routing. Furthermore, the federated  DDNet detectors, especially  the FedGS-DDNet detector, can significantly reduce the transmission overhead while protecting the data privacy and maintain satisfactory accuracy.


\linespread{1.24}
\bibliographystyle{IEEEbib}
\bibliography{References}

\end{document}